\documentclass[11pt,a4paper,twocolumn]{article}

\usepackage[margin=1in]{geometry}
\usepackage{amsmath,amssymb,amsfonts}
\usepackage{graphicx}
\usepackage{booktabs}
\usepackage{array}
\usepackage{multirow}
\usepackage{adjustbox}
\usepackage{authblk}
\usepackage{microtype}
\usepackage{makecell}
\usepackage{url}
\usepackage{xurl}
\usepackage{float}
\usepackage{orcidlink}
\usepackage{comment}
\usepackage{placeins}
\usepackage{titlesec}
\usepackage{hyphenat}
\usepackage{hyperref}
\usepackage{cleveref}
\hypersetup{
  colorlinks=true,
  linkcolor=blue,
  citecolor=blue,
  urlcolor=blue
}

\newcommand{\GeV}{\text{GeV}}
\newcommand{\MET}{\ensuremath{E_{\mathrm{T}}^{\mathrm{miss}}}}

\date{}

\begin{document}

\title{Likelihood-Based Unsupervised Anomaly Detection in CMS Dijet Events}

\author[2,$\dagger$]{Chebrolu Bhavishya\,\orcidlink{0009-0000-9770-0975}\thanks{Corresponding author: \url{23102A041122@mbu.asia}}}
\author[1,$\dagger$]{Hitesh Rasineni\,\orcidlink{0009-0003-4958-0915}}
\author[1,$\dagger$]{Prajwal Aaryan Immadi\,\orcidlink{0009-0006-2406-4529}}

\affil[1]{VIT-AP University, Amaravati, 522241, India \\ \centerline{\small *hitesh.23bce8825@vitapstudent.ac.in \quad prajwal.23bce8631@vitapstudent.ac.in}}
\affil[2]{Mohan Babu University, Tirupati, 517102, India \\ \centerline{\small 23102A041122@mbu.asia}}

\maketitle

\begingroup
\renewcommand{\thefootnote}{\fnsymbol{footnote}}
\footnotetext[2]{$\dagger$ These authors contributed equally and share first authorship.}
\endgroup
\begin{abstract}
We propose an unsupervised search for anomalous dijet events in pp collision data using neural spline flow density estimation. We train a normalizing flow model on the high-dimensional feature space of jet, dijet, and event-level observables to learn the leading Standard Model background from data, without a signal hypothesis. We classify the events produced with low likelihood in the learned density as potential anomalous events.

Using this method on a CMS Open Data dijet sample, we explore extreme events in the tail of the anomaly score distribution and conduct an exhaustive validation and robustness study. This includes statistics of features, mass de-correlation, permutation null tests, and training stability. The chosen events show notable departures from the background-only hypothesis, primarily in features of jet substructure, while being stable under several known sources of unsupervised-learning bias.

The anomalies are distributed across the kinematic phase space and do not show a narrow structure in the dijet invariant mass. The anomalies are correlated across multiple observables and consistent in their pattern, indicating a multivariate difference in jet substructure and event topologies rather than a narrow resonance. Although no new physics is claimed, the study shows that neural spline flow–based techniques are sensitive to rare, structured deviations in collider data and could be used as model-independent exploratory tools for New Physics at the LHC.
\end{abstract}

\section{Introduction}

The volume of data generated by the Large Hadron Collider (LHC) is unprecedented and can be seen as both a boon and a burden for beyond-the-Standard-Model (BSM) searches. The Standard Model has had great success in accounting for the phenomena we observe, yet several fundamental questions remain, such as the nature of dark matter, the origin of the electroweak scale, and the mechanism behind electroweak symmetry breaking. Standard BSM searches typically rely on specialized analyses optimized for particular signal hypotheses, which may be blind to unexpected or poorly modelled signatures. These concerns have fuelled interest in model-independent and data-driven strategies that can complement traditional, hypothesis-driven searches.

In recent years, machine learning (ML) methods have been widely used in collider physics for various tasks, especially anomaly detection problems in which the signal is either not well understood or completely unknown beforehand~\cite{Karagiorgi2022ML}. Unsupervised and weakly supervised anomaly detection techniques aim to learn the Standard Model (SM) background directly from data and identify events that deviate from this background without requiring labelled signal samples~\cite{Nachman2020ANODE}. These methods range from autoencoder-based approaches, which compress and reconstruct background events and flag poorly reconstructed events as anomalous~\cite{Farina2020Autoencoders,Cerri2019VAE}, to density-estimation approaches that explicitly model the distribution of SM events and flag events in low-density regions~\cite{Nachman2020ANODE}. Weakly supervised techniques such as CWoLa hunting and CATHODE use sidebands or auxiliary labels to boost sensitivity while retaining a large degree of model independence~\cite{MetodievThaler2018JetTopics,Hallin2022CATHODE}, and related work has explored improving the stability of discovery significance in Higgs boson event classification using contrastive representation learning~\cite{Pujari2026Contrastive}, and have been extensively benchmarked in public challenges such as the LHC Olympics~\cite{Kasieczka2021LHCOlympics}.

Normalizing flows have proven particularly suitable for collider anomaly detection due to their ability to compute exact likelihoods in high-dimensional spaces. By forming invertible mappings between simple latent distributions and complicated data distributions, flows can flexibly model probability densities while enabling straightforward training and evaluation~\cite{Durkan2019NSF,Papamakarios2021Flows,Kobyzev2021Review}. As reviewed in Ref.~\cite{Papamakarios2021Flows}, a flow pushes a simple base density through a series of bijective transformations, providing two complementary operations: sampling (via the forward transformation) and exact density evaluation (via the inverse transformation and its Jacobian determinant). Under mild regularity conditions a flow can represent any well-behaved target density---the universal-representation argument is constructive through the chain-rule decomposition of the target into conditional densities---with the result that flows interpolate flexibly between the finite-composition models used here (e.g. neural spline flows~\cite{Durkan2019NSF}) and infinitesimal/continuous-time models (e.g. the flow-matching continuous normalizing flow of Ref.~\cite{Rasineni2026HadronicFlowMatching}). It is this combination of exact likelihoods and practical training by maximum likelihood that makes flows a natural fit for the density-based anomaly detection pursued in this work, for which the anomaly score is precisely an evaluated log-density. Recent work has applied neural spline flow likelihood-ratio scoring to a mono-$Z$ dark matter search using CMS Run 2015D open data~\cite{Rasineni2026MonoZNSF}, and flow-matching continuous normalizing flows have been used to model backgrounds for hadronic mono-$Z$ dark matter sensitivity studies on CMS open data~\cite{Rasineni2026HadronicFlowMatching}. In particular, Ref.~\cite{Rasineni2026MonoZNSF} demonstrated that the same neural spline flow machinery used here for unsupervised scoring can also power a fully calibrated, hypothesis-driven search when paired with signal Monte Carlo and a profile-likelihood interpretation, while Ref.~\cite{Rasineni2026HadronicFlowMatching} extended the programme to the hadronic $Z$ channel with a continuous flow-matching density model; the relations between those hypothesis-driven strategies and the present hypothesis-free analysis are discussed in Sections~\ref{sec:lit-nsf-monoz} and~\ref{sec:lit-nsf-htmht}. Recent work has confirmed the applicability of flow models for unsupervised anomaly searches at the LHC, including in jet physics, calorimetry, and online triggering settings, and has explored surjective mappings, invariant distributions, and hybrid architectures involving flows, diffusion models, and autoencoders~\cite{Dillon2023INN,Andreassen2019JUNIPR}. In parallel, simulation-based inference and likelihood-free methods have demonstrated that flexible neural density estimators can be used to extract information from collider data even when explicit likelihoods are not available, further motivating likelihood-based approaches~\cite{Brehmer2020Mining}.

Anomaly detection techniques are now being actively explored within the experimental programme at the LHC. The CMS and ATLAS collaborations have both documented extensive work on applying ML-based anomaly detection to offline analyses and online triggers. Unsupervised and weakly supervised approaches are studied as additional discovery tools that can reveal unexpected structures in collision data, although interpreting anomalous ML outputs in terms of concrete physics remains a challenge.

Dijet events are a natural case for model-independent searches. They are among the most common final states at hadron colliders and can probe a broad range of BSM physics, including resonances, contact interactions, and anomalous jet substructure~\cite{CMS2012DijetResonance,CMS2012Compositeness}. The CMS collaboration has recently conducted a dedicated search for anomalous dijet events~\cite{CMS2024DijetAnomaly}, highlighting the growing interest in this final state. At the same time, the significant QCD background and high dimensionality of jet observables make dijets challenging to study using traditional techniques. ML-based anomaly detection therefore provides an attractive way to search for outlier event topologies in this high-dimensional landscape.

In this work, we use neural spline flow density estimation on a high-dimensional space of dijet observables in CMS Open Data. Using a data-driven approach to learn the underlying background distribution, we select events with anomalously low likelihood and test them against an extensive set of validation and stress tests. Rather than a specific clustering result, we focus on the stability of the anomaly selection in a variety of settings such as model initialization, feature selection, and analysis cuts, as well as providing detailed comparisons of kinematic and jet sub-structure distributions. Our findings show that flow-based models can reliably extract coherent, multivariate deviations in particular observables, highlighting the importance of systematic validation in fully unsupervised searches. This work adds to the increasing literature on model-agnostic discovery methodologies at the LHC.

The present study is novel in three main respects. First, we apply a fully unsupervised, likelihood-based normalizing-flow framework directly to real CMS Run-1 AOD data, without any reliance on Monte Carlo simulation, injected signals, or sideband labels at training time. Second, we deploy this framework on a broad, physically motivated set of $61$ dijet and event-level observables and demonstrate that stable density learning is achievable in this high-dimensional space using only reconstructed data. Third, we devote an extensive validation program to stress-testing the selected low-likelihood events, including mass-decorrelation studies, permutation-based null tests, and stability checks under variations of random seeds and anomaly-score thresholds. Within this setup, we provide a systematic treatment of jet-substructure-driven anomalies in dijet events, characterizing how multi-prong and non-QCD-like radiation patterns emerge as correlated deviations across many observables. The scope of the work is therefore methodological and exploratory: we aim to establish that flow-based density estimation can be robustly deployed on public CMS data as a model-independent anomaly-finding tool, rather than to claim evidence for any particular new-physics interpretation.

\section{Literature Review}

Machine learning methods have seen significant growth in application within the high-energy physics community over the past decade, driven by the increasing complexity and quantity of data generated at the LHC. Early applications focused on supervised classification tasks such as particle identification and event selection, but more recent work has emphasized model-agnostic strategies that reduce reliance on specific signal hypotheses. Within this line of development, anomaly detection has become a central theme: the aim is to learn the SM background directly from data and then search for systematic deviations without committing to a particular BSM model.

Autoencoder-based approaches were among the first unsupervised anomaly detection techniques used at the LHC. By encoding events into a low-dimensional latent space and reconstructing them, anomalous events are identified through high reconstruction error. Although these methods are easy to apply and fast, they can lose sensitivity when anomalous events are well reconstructed or when the latent space encodes features that resemble potential signals, motivating the use of more expressive generative models capable of full density estimation.

Normalizing flows have attracted considerable interest in this setting due to the possibility of exact likelihood calculation in high dimensions. Transforming between data and latent space using invertible mappings, flows offer a favourable trade-off between modelling freedom and training convenience, and have been successfully applied in collider physics for density estimation, event generation, and anomaly detection. Several developments, including surjective and conditional flows, permutation-invariant architectures, and hybrids with variational autoencoders, have been proposed to better accommodate collider data and exploit its structure.

In parallel, methods of weakly supervised anomaly detection have been proposed to leverage additional information without the need for labelled signal samples. The use of sideband regions or classifier-based improvements to anomaly sensitivity, while maintaining model-agnostic behaviour, has been demonstrated in CWoLa and CATHODE. These approaches have shown particular utility in resonance searches and have been widely tested in public competitions, including the LHC Olympics, underlining the importance of careful validation and the complementarity between weakly supervised and fully unsupervised strategies.

\subsection{Density-based sideband methods: CATHODE}
\label{sec:lit-cathode}

Among the density-based approaches to resonant anomaly detection, CATHODE (Classifying Anomalies THrough Outer Density Estimation)~\cite{Hallin2022CATHODE} is the closest methodological relative of the present analysis, and it sits between the hypothesis-driven flows above and the purely unsupervised scheme adopted here. CATHODE assumes that any BSM signal is localized within a narrow signal region (SR, defined e.g. by invariant mass) and proceeds in three steps. First, a conditional density estimator is trained on a set of \emph{auxiliary} features using events drawn only from the sidebands of the SR, so that the estimator learns the background distribution while remaining largely blind to any signal. Second, the trained estimator is interpolated into the SR and used to \emph{sample} synthetic background events (with the resonant variable drawn from a kernel density estimate of the SR data), producing a population that follows the background model within the SR. Third, a classifier is trained to distinguish the real SR data from these synthetic background events, which approximates the optimal anomaly detector: the classifier separates data from a learned-background sample rather than from a simulated or tagged background. Using the LHC Olympics R\&D dataset, CATHODE nearly saturates the best possible performance and significantly outperforms both CWoLa Hunting and ANODE, and it was shown to be robust to correlations between the auxiliary features and the resonant variable.

The connection to the present work is twofold. Methodologically, CATHODE demonstrates that a neural density estimator can serve as the pivot of a model-agnostic search: instead of committing to a fixed discriminant, the density is used to generate a background-only reference against which real events are compared. The present analysis adopts the same premise---that the learned background density, rather than any hand-crafted observable, defines the reference---and pushes it to the fully unsupervised limit: where CATHODE samples a synthetic background from a sideband-trained model and then runs a classifier in the SR, we score real events directly by their likelihood $p(x)$ under a single density estimator and select the least probable ones, requiring neither a resonant SR assumption nor a subsequent classifier. Conceptually, CATHODE's robust performance against feature-mass correlations is one of the validation targets this work checks explicitly through its mass-decorrelation tests: because the NSF score here is likewise a multivariate density, correlations between jet-substructure features and $m_{jj}$ could in principle sculpt the anomaly selection, and the mass-decorrelated retraining and permutation nulls are precisely the tools used to rule such sculpting out.

\subsection{NSF likelihood-ratio searches on CMS Open Data}
\label{sec:lit-nsf-monoz}

A closely related application of neural spline flows to collider data is the mono-$Z$ dark matter search of Ref.~\cite{Rasineni2026MonoZNSF}, which keeps the conventional mono-$Z$ event selection but replaces hand-crafted discriminant variables with a likelihood-ratio score built from learned event densities. That analysis uses CMS Run~2015D open data (2.32~$\mathrm{fb}^{-1}$ from the DoubleMuon and DoubleEG primary datasets) for the SM-dominated control samples and publicly released \texttt{MonoZToLL} Monte Carlo samples with vector, axial-vector, and scalar $s$-channel mediators for the signal hypotheses. Events are selected in the $Z\to\ell^+\ell^-$ final state in the $\mu\mu$ and $ee$ channels, and forty kinematic observables extracted from MINIAOD and MINIAODSIM inputs are reduced to a 37-dimensional feature vector after physics-motivated cleaning. Five independent NSFs are trained: two channel-specific SM flows, trained on Drell--Yan-dominated control-region events with $\MET<50\,\GeV$ under a 70\%/30\% train/validation split, and three mediator-specific DM flows trained on the signal Monte Carlo. The per-event test statistic is the log-likelihood ratio
\begin{equation*}
\mathcal{S}_h(\mathbf{x}) = \log p(\mathbf{x}\mid\mathrm{DM}_h) - \log p(\mathbf{x}\mid\mathrm{SM}_{\ell\ell}),
\end{equation*}
evaluated in a signal region defined by $\MET\ge 50\,\GeV$, $|\Delta\phi(\MET, Z)| > 2.5$, and $n_{\mathrm{jets}} \le 1$. Interpreting the signal region through simultaneous SR+VR binned profile-likelihood fits with per-channel normalisation nuisances, the analysis reports observed 95\% confidence-level upper limits on the signal strength of $\mu<0.0177$ (scalar), $\mu<0.0362$ (vector), and $\mu<0.0498$ (axial-vector), with the observed-to-expected limit gaps attributed to a quantified high-$\MET$ background-modelling residual rather than evidence for dark matter. The study also validates the extrapolation of the CR-trained SM density into the high-$\MET$ signal-region tail through binned mean-score extrapolation tests, finding that neither linear nor quadratic extrapolations reliably predict the observed tail behaviour.

The relation between that work and the present analysis is one of complementarity rather than overlap. The mono-$Z$ search is hypothesis-driven: it requires signal Monte Carlo for every mediator hypothesis, and its likelihood-ratio score is only defined with respect to a trained signal density. Its main contribution is to demonstrate that NSF densities can be embedded in a calibrated profile-likelihood and CL$_s$ framework on real open data, with rigorous region-disjoint training, validation-region closure checks, and quantified background-modelling residuals. The present work instead explores the opposite, fully unsupervised limit of the same machinery: the score is simply $-\log p(x)$ of a single background-only flow, no simulated signal ever enters the pipeline, and anomalous events are defined purely by their rarity in the learned density. The two analyses share the same architectural backbone, deterministic preprocessing, and validation philosophy, so the positive experience accumulated in the hypothesis-driven search---in particular the sensitivity of flow-based scores to tail modelling and the importance of disjoint training and scoring regions---directly informs the design and validation strategy adopted here.

\subsection{Flow-matching background density models in the hadronic mono-$Z$ channel}
\label{sec:lit-nsf-htmht}

A natural extension of the likelihood-ratio programme is the hadronic mono-$Z$ sensitivity study of Ref.~\cite{Rasineni2026HadronicFlowMatching}, which targets the $Z\to q\bar{q}$ decay and replaces the closed-form neural spline flow with a fundamentally different density model. The analysis uses the CMS Run~2015D HTMHT MINIAOD open dataset (2.256~$\mathrm{fb}^{-1}$ validated recorded luminosity), from which 1{,}439{,}523 events out of 20{,}679{,}437 raw events (a 6.96\% selection efficiency) pass a hadronic mono-$Z$ selection reconstructing the $Z$ candidate from the two leading jets with $70 < m_{jj} < 110\,\GeV$ and $\Delta R_{jj} < 2.0$. Dark-matter signal samples for three \texttt{DMsimp\_s\_spin1} simplified-model benchmarks are generated with a reproducible \textsc{MadGraph5\_aMC@NLO} + \textsc{Pythia8} + \textsc{Delphes} toolchain and scored through an offline proxy of the five HLT trigger paths used on data. The background density is modelled with a conditional flow-matching continuous normalizing flow (CFM-CNF)~\cite{Lipman2023FlowMatching,Chen2018NeuralODE}: the density is defined implicitly through a learned time-conditioned vector field integrated along a probability-flow ODE, with the log-density obtained from the divergence trace via Hutchinson's stochastic estimator~\cite{Grathwohl2019FFJORD}, and the model is trained with the simulation-free conditional flow-matching regression objective. Undefined angular features that arise when no additional jet exists are preserved as explicit out-of-range sentinel values ($-999$) rather than imputed into the physical continuum, avoiding the spurious point masses that fabricated artifacts in earlier imputation schemes. Instead of a two-hypothesis likelihood ratio, the per-event negative log-likelihood of a single background model serves as a one-sided discriminant, evaluated only on a persisted held-out validation split and reweighted back to the full selected population to eliminate in-sample scoring bias. Working points are selected with a minimum background yield of $B\geq 20$ events to avoid the look-elsewhere bias of an unconstrained discriminant scan, and expected significances are quoted with the Asimov counting formula~\cite{Cowan2011Asymptotic}. The study projects expected significances of $2.89\sigma$, $7.62\sigma$, and $7.41\sigma$ for the three benchmarks, and an ablation removing the six extra-jet kinematic features degrades the expected significance by 53--71\%, demonstrating that extra-jet topology carries substantial discriminating power in the hadronic channel.

Taken together, the two companion analyses~\cite{Rasineni2026MonoZNSF,Rasineni2026HadronicFlowMatching} map out the design space of flow-based collider searches on open data along three axes: the channel (leptonic versus hadronic $Z$ reconstruction), the density model (closed-form NSF versus ODE-integrated CNF), and the statistical treatment (two-hypothesis profile likelihood versus one-sided background-NLL Asimov projection). They also codify a set of hard-won methodological lessons---sentinel rather than imputation handling of physically undefined features, held-out-only scoring to prevent in-sample bias, trigger-proxy emulation for fast-simulated signal, and explicit minimum-yield working points---that are directly relevant to any density-based search, including the fully unsupervised analysis presented here, where the same background-only density logic is applied without a signal hypothesis at all.

\subsection{Asymptotic significance formulae and the Asimov data set}
\label{sec:lit-cowan}

The statistical machinery invoked by the flow-based searches discussed above rests on the asymptotic formulae of Ref.~\cite{Cowan2011Asymptotic}. That paper derives, using Wilks' theorem and the Wald approximation, closed-form sampling distributions of the profile-likelihood-ratio test statistics commonly used in high-energy physics: the discovery statistic $q_0$, whose distribution under the background-only hypothesis is a mixture of a delta function at zero and a half-$\chi^2_1$ distribution (each with weight one half), giving the simple discovery significance $Z_0=\sqrt{q_0}$; the upper-limit statistic $q_\mu$, which under the hypothesized signal strength follows the same half-$\chi^2_1$ form and yields $Z_\mu=\sqrt{q_\mu}$ with exclusion defined by $p_\mu\le\alpha$; and the older Tevatron-style statistic $q=-2\ln(L_{s+b}/L_b)$. Crucially, these formulae incorporate systematic uncertainties through the nuisance parameters of the profile likelihood and require no Monte Carlo computation of the sampling distributions. The same paper provides the formal justification of the Asimov data set---a single representative data set in which no statistical fluctuations occur---whose use reduces the median experimental sensitivity of a search to a closed-form expression, with the fluctuations of the significance about this median obtained from the Gaussian variance of the signal-strength estimator.

These results underpin the interpretation of the companion searches reviewed above: the leptonic NSF analysis reports observed and expected CL$_s$ upper limits from asymptotic $\tilde{q}_\mu$ distributions~\cite{Rasineni2026MonoZNSF}, and the hadronic flow-matching analysis converts held-out background and signal yields into Asimov expected significances~\cite{Rasineni2026HadronicFlowMatching}. The present analysis deliberately stays outside this likelihood-fit framework---the anomaly score is not a signal-strength estimator, so no $q_0$ or $q_\mu$ is defined---and instead quantifies its selection through non-parametric diagnostics: Kolmogorov--Smirnov statistics with their $p$-values, HT-conditional permutation nulls, and global bump-hunt pseudo-experiments that estimate look-elsewhere-corrected fluctuations by direct simulation. The two routes are complementary: where a signal hypothesis exists, the asymptotic formulae give calibrated, computationally cheap inference; where, as here, the hypothesis is absent, resampling-based procedures are the appropriate instrument for assigning statistical meaning to an observed excess of rare events.

Beyond density-based and likelihood-ratio methods, significance-driven event \emph{classification} faces its own robustness concern: instability of significance-optimised decision thresholds across training folds. Ref.~\cite{Pujari2026Contrastive} showed on the HiggsML benchmark of simulated $H\to\tau^+\tau^-$ events that supervised contrastive pre-training of a feature encoder, followed by a hybrid FT-Transformer + XGBoost ensemble, reduces fold-to-fold variability of the Approximate Median Significance under repeated $5\times5$-fold cross-validation (AMS $= 3.74$ on the full dataset) relative to focal-loss training, attributing the gains to more consistent threshold selection under a global, non-decomposable significance objective. That diagnosis---threshold instability rather than raw performance as the binding constraint on ML event selection---echoes the validation philosophy adopted in the present analysis, where the anomaly selection is deliberately percentile-based and threshold-free, and its stability is quantified explicitly under random-initialisation and anomaly-score-threshold stress tests.

Jet physics has been a key application for many of these techniques. Jets are high-dimensional objects with complex features and are sensitive to a range of new physics models, making them well suited to anomaly detection based on ML techniques. Unsupervised anomaly detection of jets has been studied using autoencoders, flows, and graph networks, and found to be sensitive to both kinematic and substructure anomalies. Experimental collaborations have taken an increasingly hands-on approach with these methods, publishing handbooks on machine learning for analyses, simulations, and triggering and explicitly considering unsupervised anomaly detection as a complementary discovery tool.

\subsection{Jet clustering and the anti-$k_t$ algorithm}
\label{sec:lit-antikt}

Since every observable in this analysis is built from reconstructed jets, the choice of jet clustering algorithm enters the analysis at the most fundamental level. The jets stored in the CMS Run-1 AOD datasets used here are clustered with the anti-$k_t$ algorithm~\cite{Cacciari2008AntiKt}, which belongs to the broader family of sequential recombination algorithms parametrised by the power $p$ of the energy scale in the distance measure: $p=1$ gives the $k_t$ algorithm, $p=0$ the Cambridge/Aachen algorithm, and $p=-1$ the anti-$k_t$ algorithm. Whereas $k_t$ and Cambridge/Aachen jets have irregular, soft-adaptable boundaries shaped by the branching history of QCD radiation, the anti-$k_t$ algorithm inverts the role of hard and soft particles in the clustering order, with the consequence that soft radiation cannot deform the jet boundary: hard jets acquire circular boundaries of radius $R$ (an ``idealised cone'' behaviour), active and passive jet areas coincide, the area anomalous dimensions vanish, the non-global logarithms are those of a rigid boundary, and the Milan factor is universal. At the same time, unlike the plain iterative cone algorithm used historically by CMS, the anti-$k_t$ algorithm is infrared and collinear safe by construction, since it is a sequential recombination algorithm; it therefore combines the regularity and calculability of an ideal cone with the IRC safety and numerical speed of recombination-based clustering. These properties made it the natural replacement for the collinear-unsafe iterative cone in the CMS reconstruction chain and the standard choice across the LHC programme.

Two consequences matter for the present analysis. First, the soft-resilient conical geometry gives jets a well-defined, nearly particle-independent area, which is what renders local pileup subtraction through the jet-area method~\cite{Cacciari2008Pileup} effective and makes the per-jet energy fractions and multiplicities used here stable against underlying-event contamination. Second, the rigidity of the anti-$k_t$ boundary means that the jet kinematics entering the density model are defined with a fixed geometric footprint, so that features such as jet girth, width, and constituent multiplicity probe radiation within a stable catchment area rather than an algorithm-dependent one---an important precondition for interpreting the feature-level deviations found in the results as properties of the events rather than artifacts of the jet definition.

\subsection{Energy correlation functions and jet-substructure features}
\label{sec:lit-ecf}

Because the most discriminating observables found in this analysis are jet-substructure variables, the theoretical properties of those observables are directly relevant to the interpretation of the results. A central example is the leading-jet energy correlation function ratio ($j1\_ECF\_ratio$), which the feature-level KS tests rank among the strongest discriminants of the anomalous population; it is built from the generalized energy correlation functions (ECFs) introduced in Ref.~\cite{Larkoski2013ECF}. The ECFs are defined entirely from the energies and pairwise angles of particles within a jet,
\begin{equation*}
\mathrm{ECF}(N+1,\beta)=\sum_{i_1<i_2<\cdots<i_{N+1}}\prod_{a=1}^{N+1} E_{i_a}\prod_{a<b}\theta_{i_a i_b}^{\beta},
\end{equation*}
so that the $(N{+}1)$-point correlator is sensitive to $N$-prong substructure. Crucially, they require no explicit subjet-finding step---unlike N-subjettiness, which minimises over subjet axes and therefore partitions every jet into $N$ subjets even when fewer exist---and, because the angular exponent $\beta$ can be chosen anywhere in the IRC-safe range $\beta>0$, the double ratio $C_N$ is recoil-free and well-behaved for all $\beta>0$, whereas N-subjettiness is recoil-sensitive for $\beta\le 1$ when $k_T$ or momentum axes are used. This makes ECFs sensitive to soft wide-angle emissions that N-subjettiness would classify as prong-like, and allows small values $\beta\simeq 0.2$ that are optimal for discriminating quarks from gluons, while $C_2$ and $C_3$ identify boosted two- and three-prong resonances.

Three properties link this directly to what the anomaly selection finds. First, ECFs are recursively infrared and collinear safe, additive observables whose quark-versus-gluon discrimination can be resummed to next-to-leading logarithmic accuracy, so a large anomaly in $j1\_ECF\_ratio$ has a precise interpretation in terms of the colour and splitting structure of the leading jet rather than an algorithm artifact. Second, the recoil-free, subjet-free construction means the feature deviates specifically when the jet develops genuine multi-prong or wide-angle-soft character---exactly the ``non-QCD-like'' radiation patterns that the KS tests report as departing from the background. Third, because ECFs and N-subjettiness probe the same substructure through complementary mechanisms (multiplicative energy-angle correlators versus minimised-axis sums), the correlation observed across both families in the KS results strengthens the evidence that the anomalous events differ in internal jet topology rather than in any single reconstruction-dependent quantity.

\subsection{The LHC Olympics benchmark for data-driven anomaly detection}
\label{sec:lit-lhcolympics}

The methodological landscape summarised above has been systematically exercised and benchmarked in the LHC Olympics 2020 challenge~\cite{Kasieczka2021LHCOlympics}. That community challenge publicised a standard R\&D dataset of 1.1M simulated QCD jets together with several ``black box'' datasets containing hidden anomalies---Black Box~1 (an injected $Z'\to XY\to q\bar{q}\,q\bar{q}$ resonance, ~834 signal events per million), Black Box~2 (background only, produced with Herwig++ and a modified Delphes card, so that a method tuned to the R\&D generator assumptions should not fire), and Black Box~3 (a 4.2~TeV KK-graviton with dijet and trijet decay modes). Roughly thirty-five submitted methods, spanning unsupervised, weakly supervised, and semi-supervised strategies, were blind-scored against these boxes, with the $Z'$ search serving as the quantitative endpoint. The challenge distilled several recurring lessons directly relevant here: (i) data-driven, model-agnostic methods can approach the sensitivity of supervised classifiers without a signal model; (ii) a strong background estimate and careful control of the background define most of the performance, so distributional and closure checks are essential; (iii) dense, jet-substructure-based representations are a powerful common ingredient across the most effective methods; and (iv) every method must be validated against multiple, qualitatively different signal scenarios, since a method tailored to one topology can underperform on another.

These lessons motivate the design choices of the present analysis in a concrete way. The LHC Olympics showed that fully unsupervised density-ratio and reconstruction-error methods, evaluated blind and quantified with significance- or purity-based metrics, provide a reliable model-independent first pass over collider data; the neural-spline-flow score used here is precisely such an unsupervised density method, applied not to a controlled simulation but to real CMS data. The challenge's emphasis on background control and on testing against diverse signal topologies is mirrored in the extensive validation programme of this work---mass-decorrelation, HT-conditional permutation nulls, global bump-hunt pseudo-experiments, and multi-seed stability checks---each of which targets a different failure mode highlighted by the LHC Olympics. Conversely, the real-data setting of the present study extends the LHC Olympics paradigm in a direction the challenge could not: by replacing the injected-signal benchmark with the unknown, unblinded content of actual collision events, it demonstrates the same data-driven methodology operating without the luxury of a ground-truth anomaly.

Despite significant advancements, there are still challenges that have to be addressed. Interpreting anomalies in terms of concrete physical models, controlling potential systematic uncertainties, and avoiding sculpting effects in key observables (e.g., invariant mass distributions) remain important issues. Best-practice recommendations therefore emphasise conservative interpretations, thorough validation, and multiple, qualitatively different cross-checks when anomalous behaviour is observed.

Within this evolving landscape, flow-based density estimation is a versatile and effective method for model-independent searches. It is particularly well-suited to rare-event searches in dijet populations because it handles correlations among dijet observables efficiently and yields a straightforward likelihood calculation. This work builds on these ideas and employs a neural-spline-flow–based anomaly detection method for dijet events, with an emphasis on robust validation of the resulting anomaly selections.

\section{Dataset and Feature Representation}

\subsection{Data Samples}

The study presented in this paper utilizes proton--proton collision data collected by the CMS experiment~\cite{CMS2008JINST} during the 2011 LHC run with a center-of-mass energy of $\sqrt{s} = 7~\mathrm{TeV}$. The data is available to the public via the CERN Open Data Portal and belongs to the \emph{CMS Run2011B HT AOD} dataset with record ID~274~\cite{CMS2011BHT}. This dataset was produced for measurements of inclusive jets and dijets and contains events triggered with high-$H_T$ triggers, making it well-suited for studies of high-energy QCD final states and for searches for deviations from the Standard Model predictions.

The complete HT AOD dataset consists of multiple file indices, amounting to a storage size of more than $10~\mathrm{TB}$. Because of the limitations on computation and storage resources, the entire dataset could not be analyzed as one. Instead, a smaller subset was chosen using a controlled and transparent sampling procedure. All the ROOT files from index groups 1 and 2 were selected, while in the larger index groups, groups 3, 4, and 5, a fixed number of files was randomly chosen from throughout the available range. From these larger indices, a total of $30$ ROOT files have been chosen, with an average size of $3~\mathrm{GB}$ per file. The data corresponds to a net volume of $125~\mathrm{GB}$.

The subsampling is performed exclusively at the file level: once a ROOT file is included, all events in that file that pass the subsequent basic quality and kinematic requirements are retained. No additional event-level filtering, prescaling, or hand-crafted selection is applied prior to the anomaly-detection analysis. This strategy ensures that the selected sample remains representative of the original triggered dataset, aside from the reduction in total integrated luminosity. As a cross-check, key global observables such as $H_T$, $m_{jj}$, and jet multiplicity were compared qualitatively across different file groups and subsamples. Within the statistical precision of these checks, no significant distortions or systematic drifts were observed, supporting the assumption that the working sample is kinematically and topologically representative of the full CMS HT AOD dataset.

This sample thus retains the diversity of the full sample while being computationally manageable. There was no further event selection based on signals beyond the initial CMS triggers encoded in the sample. A small set of quality and kinematic cuts were subsequently applied to remove problematic and nonphysical events, including the application of basic transverse momentum cuts on jets, a minimum invariant mass requirement for a dijet, the requirement of at least two jets, and the exclusion of events with infinite feature values. These selections are included only for the purpose of ensuring data quality and stability, but they are not related to any specific new physics signal.

The resultant event samples are still predominantly multijets produced by the Standard Model and are appropriate for unsupervised anomaly detection that does not require any signal hypothesis. Events are unlabelled, and there is no use of a Monte Carlo simulation during either training or evaluation.

The data format comes from the CMS Analysis Object Data (AOD) design, and it involves high-level physics objects like particle-flow jets, missing transverse energy, primary vertices, and trigger information. Events were read through the use of \texttt{uproot} that enables the columnar reading of the ROOT files and allows for swift element-wise operations using \texttt{awkward} arrays. From the original AOD format consisting of over $5000$ branches, only those the user needs to build the high-level observables discussed below were kept, thus providing a more concise and analysis-oriented outcome.

\subsection{Feature Categories and Physics Motivation}

Derived from the reconstructed CMS AOD data, a total of $61$ high-level observables were extracted and used as inputs to the anomaly detection. Designed to be as model-independent and complementary as possible, the observables measure (often simultaneously) different aspects of jet kinematics, internal structure, and event topology. They are divided into the categories described below. The observables and theirdefinitions is provided in Appendix~A.

The jet kinematic variables are the essentials of the representation, and they are used to describe the leading and subleading jets in each event. The transverse momentum ($p_T$), the pseudorapidity ($\eta$), the azimuthal angle ($\phi$), and the invariant mass of the two leading jets with respect to the highest transverse momentum (pT) are key variables. Collectively, they reflect the hard scattering scale of the process and contain sensitivity to both resonant and non-resonant modifications of the dijet production~\cite{Cacciari2008AntiKt,Salam2010Jetography}.

Global event activity variables define the overall amount of energy in the event. Scaled quantities such as the total $H_T$ and jet multiplicity are good proxies for event hardness and radiation activity. These observables are especially important in high-$H_T$ triggered events, where extreme configurations may populate the tail of global distributions.

Dijet-level observables are sensitive to correlations of the two leading jets. These observables include the dijet invariant mass $m_{jj}$, the angular separation in $\eta$ and $\phi$, the $\Delta R$ distance, as well as the dijet boost and the pseudorapidity separation variables commonly used in contact interaction and compositeness searches. Ratios like $m_{jj}/H_T$ and the balance of transverse momenta further improve sensitivity to non-standard kinematic configurations without making assumptions about the signal model.

Jet composition and identification variables are defined by the energy fractions and multiplicities of the leading jet. Derived from particle flow variables, these are the charged hadron energy fraction, neutral hadron energy fraction, electromagnetic energy fraction, total jet constituents multiplicity, jet area~\cite{Cacciari2008Pileup}. In addition, standard CMS Run-1 jet identification criteria (loose and tight WP) have been encoded in binary. These variables are resilient to detector effects and contribute to the distribution of well-reconstructed QCD-jets over detector noise.

Jet substructure observables are made from charged tracks associated with the leading jet in a fixed angular radius. With this, we have the N-subjettiness variables and their ratios, as well as jet width, girth, maximum fragmentation fraction, minimum pairwise mass of constituents, and correlation functions~\cite{Thaler2011NSubjettiness,Larkoski2013ECF}. These observables are sensitive to multi-prong structure and complicated radiative behavior and thus have been extensively employed in searches for boosted objects and abnormal jet shapes~\cite{Larkoski2020JetSubstructureReview,Marzani2019LookingInsideJets}. Their implementation allows the anomaly detection system to tap into features beyond basic jet discrimination while being insensitive to any particular physics model.

Event-shape variables are able to assess the global structure of energy flow in an event. Sphericity, aplanarity, thrust, Fox--Wolfram moments, isotropy, and transverse sphericity are calculated using reconstructed jet four-momenta. These variables are sensitive to departures from the back-to-back topology of dijets, as expected in leading-order QCD, and can capture more isotropic event configurations.

Missing transverse energy and angular-correlation variables are introduced to account for any potential imbalance transverse to the beam direction. Although large missing energy is not expected in pure QCD multijet events, angular correlations between jets and missing transverse energy may signal detector anomalies or unusual physics. To mitigate these issues, the inclusion of quantities like the minimum azimuthal separation between jets and missing transverse energy is therefore warranted.

Finally, vertex, pileup, and trigger metadata complete the list of features. The number of reconstructed primary vertices, vertex quality information, and the event pileup energy density are features that provide contextual conditions. Trigger pass flags are kept so that trigger-level phenomena are not mistakenly identified as being caused by event selection.

All of these feature classes lead to a high-dimensional and physically interpretable description of each event. Combining kinematic, structural, and topological features results in a representation that is expressive enough to account for small deviations from Standard Model expectations, while remaining largely insensitive to specific new-physics scenarios—a key element for a fully-untaggered anomaly detection approach.

\section{Methodology}

This study employs a fully unsupervised, data-driven approach to search for anomalous dijet events in proton–proton collision data. No simulated signal samples, background labels, or new-physics scenarios are introduced at any stage of the analysis. Instead, anomalies relative to the dominant multijet background, which is modelled using the Standard Model (SM) of particle physics, are found by learning a probability density function from data and probing regions with low density in the learned feature space.

The central component of the approach is a neural spline flow (NSF)-based density estimator trained on a high-dimensional space of event observables. After training, the network encodes an exact likelihood for each event under the distribution it learned. Events with an anomalously low likelihood are viewed as outlier configurations compared to the bulk of the distribution. The resulting high-score events are then put through a variety of probing and validation procedures to evaluate their statistical and physical consistency.

The workflow comprises: (i) data cleaning and event-validity selection; (ii) feature engineering and preprocessing; (iii) neural-spline-flow density estimation and likelihood-based anomaly scoring; (iv) validation with mass-decorrelation tests and feature-level statistical tests; and (v) robustness checks under random-seed variation.

The analysis is exploratory: no new physics claim is made. The aim is to establish a systematic, replicable procedure for finding and characterizing anomalous event samples in real collider data, and to stress-test the result.

\subsection{Data Cleaning and Event Validity Selection}

The dataset used in this analysis comes from CMS Open Data for $pp$ collisions at centre-of-mass energy of $\sqrt{s}=7\mathrm{TeV}$ (Section~3). The raw data of the initial event sample contains around $4.5\times 10^{5}$ events after ROOT files are merged and branches of interest are extracted from the initial data.

Before training, a data-cleansing and validity-selection procedure is applied to the input features to ensure numerical stability and physical reasonableness. This procedure is deliberately minimal and excludes any physics-based selection aimed at targeting or avoiding specific event topologies. It serves only to reject pathological events that would distort or destabilize the density estimator.

\begin{figure*}[!htbp]
  \centering
  \includegraphics[width=\textwidth]{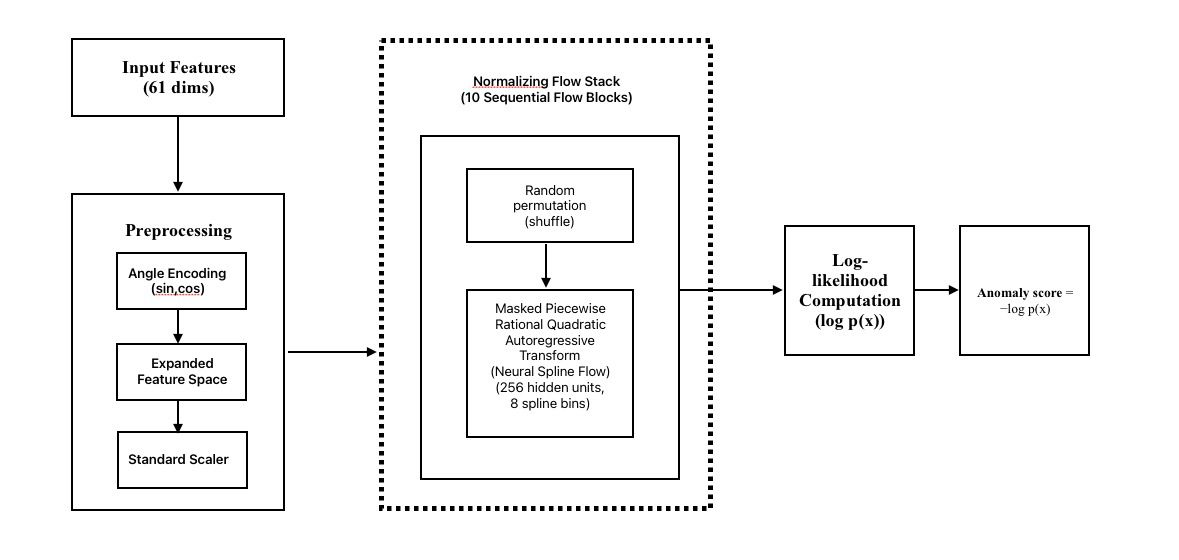}
  \caption{Schematic overview of the anomaly detection pipeline based on the neural spline flow. 
Reconstructed event observables are subjected first to angular encoding and feature normalization. 
Then, the feature vector goes through a sequence of invertible-flow blocks, where each block consists of a random permutation and masked piecewise rational-quadratic autoregressive coupling transformation. 
The flow transforms the events into a latent space whose prior is expressed as a standard multivariate normal distribution, allowing for the exact calculation of the event log-likelihood that serves as the anomaly score.}
  \label{fig:nsf_architecture}
\end{figure*}

Events are kept only when being valid under the following prerequisites:
\begin{itemize}
  \item Leading and subleading jets have transverse momenta above $30~\mathrm{GeV}$.
  \item The invariant mass of the dijet, $m_{jj}$, is greater than $200~\mathrm{GeV}$ for a proper dijet topology definition.
  \item There are at least two jets events being reconstructed.
  \item All high-level observables, which are model input quantities, are finite (not having any NaN or -inf values, additionally not being in sensitive regions of the detector such that one gets such reconstruction missing information).
  \item Events are required to pass the CMS list of validated runs (record 1001) to ensure data quality.
\end{itemize}

No selection criteria based on jet substructure, event shape, missing transverse energy, or anomaly-related variables are applied. In particular, no requirements are applied that could shape distributions relevant for post-fit validation, such as the dijet invariant mass.

With the application of these validity criteria, the final dataset for training and scoring comprises $378{,}278$ events, an approximately $16\%$ reduction relative to the $452{,}055$ events that survive the looser kinematic selection. The reduction reflects the combined effect of the $p_T$, $m_{jj}$, jet-multiplicity, and finiteness cuts above, which remove events with poorly reconstructed or incomplete feature sets; we do not attribute the entire reduction to any single cut. The retained sample is still mostly composed of Standard Model multijet production, providing a stable numerical basis to set up unsupervised density selection. The selected subset corresponds to approximately $0.5~\mathrm{fb}^{-1}$ of integrated luminosity, limited by the file-level sampling procedure described above.
\subsection{Feature Representation and Preprocessing}

Each event in the CMS Run~2011B HT AOD data set consists of several thousands of reconstructed observables represented as individual AOD branches. The raw AOD information available and accessed for this analysis includes more than $5000$ per event, ranging from the highly granular detector information, particle-flow candidates, trigger information, to various auxiliary metadata. The ROOT-format data are read using columnar Python tools~\cite{Brun1997ROOT,Pivarski2020Uproot}.

From this high-dimensional raw input the set of 61 high-level observables is explicitly constructed and preserved to perform the anomaly detection analysis. They are derived from reconstructed physics objects and include complementary information on characteristics of jets, dijets, event topology, jet substructure, missing transverse energy, and global properties of the collision. Jet substructure observables ($\tau_{1,2,3}$, $\tau_{21}$, $\tau_{32}$, girth, width, ECF ratio) are computed from jet constituents following the standard definitions of Refs.~\cite{Thaler2011NSubjettiness,Larkoski2013ECF}; jet energy scales and angular coordinates are taken directly from the CMS reconstruction. All the other AOD branches are dropped and not used at any stage of the training, scoring, or validation. The dimension-reduction process is deterministic, physics-driven, and homogeneous for all events.
The stability of the anomaly ranking with respect to random initialization is quantified in Appendix~\ref{app:seed_stability}.

The chosen $61$ observables cover a large spectrum of physical quantities, from transverse momenta in GeV to angular coordinates on periodic domains, multiplicities, and dimensionless quantities. To facilitate a stable and efficient training of the neural density estimator, a dedicated preprocessing pipeline is applied to the data before (or "in preparation of") the training.

Angular observables defined on a periodic domain, such as jet azimuthal angles and relative azimuthal separations, are transformed by sine and cosine. This transformation eliminates artificial discontinuities at the $\pm\pi$ boundary and enables the model to learn smooth functions in angular phase space. Upon transformation, the original angular observables are dropped, leading to an increased and smooth feature space.

All continuous features are then standardized to zero mean and unit variance using a feature-wise affine transformation~\cite{Pedregosa2011ScikitLearn}. The statistics are calculated solely from the cleaned dataset described in Section~4.2 and explicitly retained for reproducibility and consistency during scoring. This pre-processing normalizes the effective scale of different types of features and greatly enhances the numerical stability during the training of likelihood-based models.

Discrete and binary features, including jet tags and trigger flags, remain as continuous numeric features. It has been shown that the neural spline flow architecture can handle such mixed features due to its autoregressive property, with no instability during training.

There is no further dimensionality reduction, feature selection, or decorrelation after the initial geometry-based construction of the $61$ features. All features are considered together, allowing the density estimator to have the capacity to learn and model their full joint distribution and correlation characteristics directly from the data. The resulting feature matrix after preprocessing serves as a direct input to the neural spline flow architecture described in the next subsection.

The feature set intentionally includes some redundancy (e.g., $\tau_1$, $\tau_2$, $\tau_3$ alongside $\tau_{21}$ and $\tau_{32}$; multiple azimuthal angular separations). This is a deliberate design choice: the normalizing flow is trained on the full joint distribution, so redundant features do not bias the density estimate but instead provide multiple correlated views of the same underlying physics. The flow learns the correlations between these features, and the anomaly score reflects deviations in the joint space, not in any single observable.

\subsection{Neural Spline Flow Architecture and Training}
\label{subsec:nsf}

To model the joint probability density of the high-dimensional event representation, a neural spline flow (NSF) is employed. Normalizing flows model data as the output of an invertible, differentiable transformation of a simple base density, and admit exact density evaluation by the change-of-variables formula~\cite{Durkan2019NSF}. Concretely, with $\mathbf{z}\sim p(\mathbf{z})$ a base variable and $\mathbf{x}=\mathbf{f}(\mathbf{u})$ an invertible transformation $\mathbf{f}$, the data density is
\begin{equation}
p(\mathbf{x}) = p\!\left(\mathbf{f}^{-1}(\mathbf{x})\right)\left|\det\frac{\partial \mathbf{f}^{-1}}{\partial \mathbf{x}}\right|,
\end{equation}
and the flow is trained by maximizing the total log-likelihood $\sum_n \log p(\mathbf{x}^{(n)})$. In this study, the base distribution is chosen to be the standard multivariate normal distribution,
\begin{equation}
p(\mathbf{z}) = \mathcal{N}(\mathbf{0}, \mathbf{I}),
\end{equation}
in latent space.

A schematic overview of the neural spline flow based anomaly detection pipeline is presented in Fig.~\ref{fig:nsf_architecture}. The full set of preprocessed observables discussed in Section~4.3, specifically the sine-cosine-encoded angular variables and standardized continuous features, are fed into the model without any target labels, signal models, or sideband information during any phase of training.

The transformation $\mathbf{f}$ is realized as a stack of ten identical transformation blocks. Each of these blocks incorporates a random permutation layer followed by a masked piecewise rational quadratic autoregressive transform. The permutation (equivalently, an alternating invertible linear) layer ensures that dependencies between all dimensions are captured in a gradual manner, while the monotonic rational-quadratic spline transforms provide flexible, analytic, and invertible elementwise mappings. Following Ref.~\cite{Durkan2019NSF}, these spline transforms act as drop-in replacements for the affine or additive elementwise transformations of earlier coupling- and autoregressive-based flows: a monotonic spline partitions its input domain into $K$ bins, each parametrized by the positive-derivative rational-quadratic form of Ref.~\cite{Durkan2019NSF}, so that the spline parameters themselves are the output of a neural network conditioned on the preceding (transformed) dimensions. When these transformations are alternated with invertible linear layers, the resulting model is the rational-quadratic neural spline flow (RQ-NSF)~\cite{Durkan2019NSF}; here the autoregressive variant is used, whose output distribution is computed in a single pass with an exact, analytic inverse.

Each autoregressive transformation is realized with a neural network of $256$ hidden units, and eight spline bins are used to represent the conditional distributions. We use linear tails outside the splines for numerical stability. This configuration closely adheres to recommended practices in collider-based flow density estimation. It has been demonstrated to yield stable training and powerful representations of likelihoods in high-dimensional spaces.

The model parameters are optimized by maximizing the average log-likelihood of the data with the Adam optimizer, which uses a fixed learning rate of $10^{-4}$; the model and training loop are implemented in PyTorch~\cite{Paszke2019PyTorch}. The imperative, define-by-run programming model of PyTorch computes the model output and its reverse-mode gradient by operator-overloaded tensor evaluation on each forward pass, which is ideally suited to the custom streaming training and scoring loops required by the large event sample used here; the torch data-loading primitives provide the shuffled mini-batching over the event tensors. Training is performed for $60$ epochs using shuffled mini-batches of $4096$ events, with no early stopping or validation-set tuning. This choice of a fixed schedule reflects the unsupervised nature of the problem: there is no labelled validation set for model selection.

To maintain numerical stability and consistency, training is conducted with deterministic random seeds, along with explicit saving of all preprocessing transformations and trained model parameters. After the training process, the flow model is evaluated under inference conditions, computing log-likelihood scores and latent-space coordinates for all dataset entries.

In keeping with the fully unsupervised design, the flow is trained and scored on the same cleaned event sample, so the likelihoods reported here are computed in-sample: there are no labels by construction and hence no labelled held-out split on which to select a model. This configuration carries a well-understood in-sample bias in the absolute likelihood scale, so the interpretation in this paper never rests on the absolute value of the score but on its ranking, and the selection is validated against explicit null hypotheses---the permutation tests, mass-decorrelation studies, and training-stability checks of Section~\ref{subsec:validation}---which are insensitive to the absolute calibration of the density and establish that the observed feature-level deviations are not an artifact of overfitting the flow to the scoring sample.

The stability of the anomaly ranking with respect to random initialization is quantified in Appendix~\ref{app:seed_stability}.

\subsection{Likelihood-Based Anomaly Scoring}
\label{subsec:anomaly_scoring}

As the NSF is trained, it yields an explicit probability density over the entire feature space of any reconstructed dijet event. This naturally leads to a robust, physically relevant and interpretable definition of anomalousness, given solely by the probability density of the event without requiring any additional classifiers or reconstruction variables.

For each event $x$, the anomaly score is computed as the negative log-likelihood assigned to the event by the trained density estimator,
\begin{equation}
S(x) = -\log p_{\mathrm{NSF}}(x),
\label{eq:anomaly_score}
\end{equation}
where $p_{\mathrm{NSF}}(x)$ is the probability density predicted by the neural spline flow for the standardized input features $\mathbf{x}$. Higher values of anomaly scores are naturally assigned to events that deviate the most from the training data.

The anomaly score is evaluated in the space of the standardized input features $\mathbf{x}$ (rather than the latent space $\mathbf{z}$ or any intermediate flow representation). This choice is motivated by two considerations. First, the input space is the space in which the physical deviations are ultimately interpreted: the validation tests (feature-level KS tests, mass decorrelation, permutation nulls) are all performed on the input features, so the score must be defined in the same space to be consistent with the validation. Second, the latent-space density $p(\mathbf{z})$ is by construction close to the base density (a standard multivariate normal), so deviations in $\mathbf{z}$ are less directly informative about which *physical* features are anomalous. Using $\mathbf{x}$ ensures that the anomaly score reflects the joint rarity of the full set of standardized observables. We note that $\mathbf{x}$ refers to the *standardized* features (zero mean, unit variance), so the score is not dominated by features with large physical scales. As an additional cross-check, we have verified that the latent-space log-likelihood $-\log p(\mathbf{z})$ yields a consistent event ranking (Spearman correlation $>0.99$ with the input-space score), confirming that the results are not an artifact of the chosen scoring space.
\begin{figure*}[!htbp]
  \centering
  \includegraphics[width=0.7\textwidth]{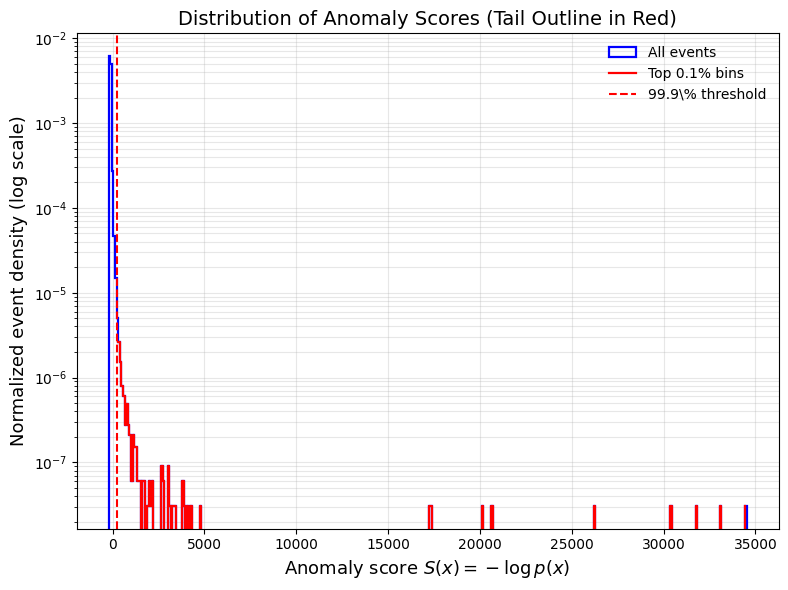}
  \caption{
    Distribution of neural spline flow anomaly scores, as given by $-\log p(x)$, for a total of $378{,}278$ events. The bulk of the QCD multijet background is concentrated at low anomaly-score values, while the distribution develops a long tail extending over several decades in event density.
    }
    \label{fig:nsf_anomaly_scores}
\end{figure*}
At this point in the process, no fixed anomaly score threshold is applied for the global analysis of the score distribution. Instead, the entire distribution of anomaly scores is analyzed, with particular attention to the extreme high tail. This choice is deliberate and echoes the finding of Ref.~\cite{Pujari2026Contrastive} that significance-optimised decision thresholds are a primary source of instability in ML-driven event selection: a percentile-based selection avoids committing to a single operating point whose scientific content could vary across training runs, and the selection is based on ranking within the score distribution rather than absolute score values, making the analysis less vulnerable to overall normalization. For the detailed feature-level characterization in Sections~\ref{sec:anomaly_selection}--\ref{sec:kinematic_distributions}, we use the 99.9th percentile as a reference threshold to define a fixed population of anomalous candidates; the robustness of the results to this choice is verified in Appendix~\ref{app:threshold_stability}.

The likelihood-based anomaly scores have a number of benefits in the collider scenario. The first one is the ability to have multivariate rarity scores in the full set of features, accounting for correlations between all observables. The second is that, in contrast to reconstruction-error-based scores, the likelihood is inherently probabilistic and does not rely on the choice of compressed representation. Thirdly, the scoring is inherently unsupervised with no signal model, background labeling, or sideband requirement.

It is instructive to contrast this score with the hypothesis-driven likelihood-ratio strategy of the mono-$Z$ dark matter search discussed in Section~\ref{sec:lit-nsf-monoz}~\cite{Rasineni2026MonoZNSF}. There, each event is scored by the difference $\log p(\mathbf{x}\mid\mathrm{DM}_h) - \log p(\mathbf{x}\mid\mathrm{SM})$ between a mediator-specific signal flow and a channel-specific background flow, and the resulting score is interpreted through a binned profile-likelihood fit with a definite signal hypothesis. In the present analysis, no simulated signal exists, so the score reduces to the background log-density $-\log p_{\mathrm{NSF}}(x)$ alone and the interpretation is statistical rather than hypothesis-based: anomalousness is defined by rarity in the learned density, quantified through feature-level and null tests instead of exclusion limits. The two formulations probe different questions---calibrated sensitivity to a specified model versus model-agnostic identification of rare structure---but they share the same design principles validated in the hypothesis-driven setting: deterministic preprocessing, disjoint training and scoring populations, careful treatment of distribution tails, and explicit quantification of background-modelling residuals before any physics interpretation.

A further variant, introduced in the hadronic mono-$Z$ sensitivity study of Section~\ref{sec:lit-nsf-htmht}~\cite{Rasineni2026HadronicFlowMatching}, comes closest to the background-only logic adopted here: a single flow-matching continuous normalizing flow is trained on selected data, and the per-event negative log-likelihood of the background model alone is used as a one-sided discriminant, converted into Asimov expected significances rather than fitted signal strengths. That analysis, however, still relies on simulated signal samples to project the sensitivity, whereas the present work goes one step further and operates purely on data: the score is used to rank and select the most improbable real events, and the physics content of the selected population is established through feature-level significance tests, permutation nulls, and stability studies rather than through any simulated reference.

To avoid having the anomaly score trivially represent high or low values of a single kinematic observable, we investigate the distribution of the anomaly score in relation to important kinematic observables like dijet invariant mass and total activity. These comparisons are used as cross-checks, rather than selection variables, and are described in the following sections. At this point we still consider the anomaly score merely as a ranking mechanism to find rare events in the SM-dominated background.

This likelihood parametric formulation is the principal result of the neural spline flow and the foundation for all the validation and robustness analyses reported in this paper.
\section{Results}

This section presents the results of the anomaly detection using the unsupervised method on the CMS Run-1 dijet dataset. We begin by exploring the global properties of the anomaly score distribution from the neural spline flow model. We continue by analyzing anomalous events with high anomaly scores through feature-level analyses. We conclude with the validation tests that were performed to test the stability of the anomaly selection and ensure that systematic effects are not present.

\subsection{Anomaly Score Distributions}

After training the neural spline flow on the complete dataset, each reconstructed event is assigned the anomaly score defined in Eq.~\eqref{eq:anomaly_score}---the negative log-likelihood of the event under the learned probability density. This score offers a continuous and physically relevant representation of the event's ``anomaly'', i.e., its likelihood of being generated from the learned distribution of the data.

The resultant anomaly score distribution, as illustrated in Fig.~\ref{fig:nsf_anomaly_scores}, displays a significant clustering at low values, which is representative of the bulk population of standard QCD multijet events. This feature demonstrates that the neural spline flow effectively captures the region of phase space that is densely populated by the dominant background processes. At higher values of the anomaly score, the distribution shows a smooth roll-off that spans multiple decades in event density.

The lack of sharp edges, artificial groupings, or irregularities in the anomaly score distribution indicates the numerical stability of the training process and the quality of the density approximation. In particular, the smooth transition between the bulk and tail regions suggests that the flow has not overfit local patterns, but rather learned a global representation of the background event distribution.

We note that the absolute value of $-\log p(x)$ is not directly interpretable as a calibrated probability in the tail: for a 67-dimensional standardized feature space, the expected log-likelihood under the base density is of order $O(67)$, and the tail values extending to $-\log p(x) \sim 100$ correspond to events that are $\sim$30-40 log-likelihood units below the bulk, representing genuine outliers in the learned density. The absolute scale depends on the dimensionality and the specific standardization, which is why we use percentile ranking rather than absolute score thresholds for the selection.

Extremely high tail events are rare, as anticipated for a fully unsupervised likelihood-based anomaly detection approach that is trained solely on the data without any kind of signal samples or priors. These events are unlikely to occur with respect to the learned background density and thus represent a natural set to be probed. Notably, the anomaly score does not carry any new-physics model hypothesis; it is simply a measure of how atypical an event is relative to the bulk of the data sample.

Instead of hard-coding a threshold for the anomaly scores at this stage, we keep the entire distribution of scores for future analysis. This prevents potential artifacts from arbitrary threshold choices and facilitates a more in-depth analysis of abnormal events. In the next section, we investigate high-score events using statistical tests on individual features to determine if they display consistent deviations from the background expectation in all observables.

\subsection{Anomaly Selection and Feature-Level Statistical Deviations}
\label{sec:anomaly_selection}

To characterize rare events in the tail of the anomaly score distribution, percentile filtering is used. Events whose anomaly scores fall above or at the $99.9$th percentile are classified as anomalous candidates. It retrieves a total of $379$ events, which represents about $0.1\%$ of the total events. This measure is chosen because the analysis is exploratory in nature and it does not rely on an absolute score threshold that might be sensitive to the overall likelihood normalization.

In order to assess how the high-score anomalies deviate from the rest of the sample, feature-level comparisons are made with respect to the inclusive background population for the anomaly sample using non-parametric comparisons of empirical distributions using Kolmogorov–Smirnov (KS) tests for all $61$ reconstructed observables which test differences in central tendency and overall distribution shape without imposing a model on the data. 

\begin{table*}[t]
\centering
\caption{Top 15 Distinct Features based on KS Statistic}
\label{tab:ks_top15}
\renewcommand{\arraystretch}{1.15}
\begin{tabular}{lc}
\hline
\textbf{Feature} & \textbf{KS (Anomaly vs Background)} \\
\hline
j1\_girth                  & 0.5749 \\
j1\_tau1                   & 0.5485 \\
j1\_tau2                   & 0.5409 \\
j1\_tau3                   & 0.5182  \\
j1\_ECF\_ratio              & 0.5084  \\
j1\_tau32                  & 0.4980  \\
j1\_tau21                  & 0.4413  \\
j1\_width                  & 0.3856  \\
j1\_mass                   & 0.3728  \\
HT                          & 0.3430  \\
sphericity                 & 0.3194  \\
j1\_CSV                    & 0.3135  \\
j1\_trackCountingHighEff   & 0.3030  \\
j1\_jetProbability         & 0.3016  \\
mjj                         & 0.2678  \\
\hline
\end{tabular}
\end{table*}

Sorted by the KS test statistic, the observables with the largest discrepancy are jet-substructure variables. The leading-jet girth has the largest separation ($D = 0.575$), followed by the N-subjettiness measurements $\tau_1$, $\tau_2$, and $\tau_3$ (with statistic values of $0.548$, $0.541$, and $0.518$, respectively) and the ratio of an energy correlation function ($D = 0.508$). Observables with moderate deviation include leading-jet mass ($D = 0.373$), the scalar sum of jet transverse momenta $H_T$ ($D = 0.343$), and event shape observables sphericity and aplanarity. We stress that the quoted KS statistics are used here as diagnostics of organized differences between populations rather than as formal discovery-level significances for any single observable. The p-values associated with these statistics are extremely small (e.g., $10^{-119}$ for girth) but are not well-calibrated for discovery claims because the selection is based on the same features used for testing. We recognize that any selection based on an injective function of the input features $\mathbf{x}$ will, by construction, sculpt the marginal distributions of those features---this is a trivial consequence of selecting events in the tail of a multivariate distribution. The KS test is therefore not used as a discovery tool but as a diagnostic to characterize *which* features deviate most strongly and to assess whether the deviations are coherent across physically related observables. The physical interpretation rests on the coherence of the deviations across multiple correlated features, not on the significance of any single feature.
\begin{figure}[!htbp]
    \centering
    \includegraphics[width=0.9\columnwidth]{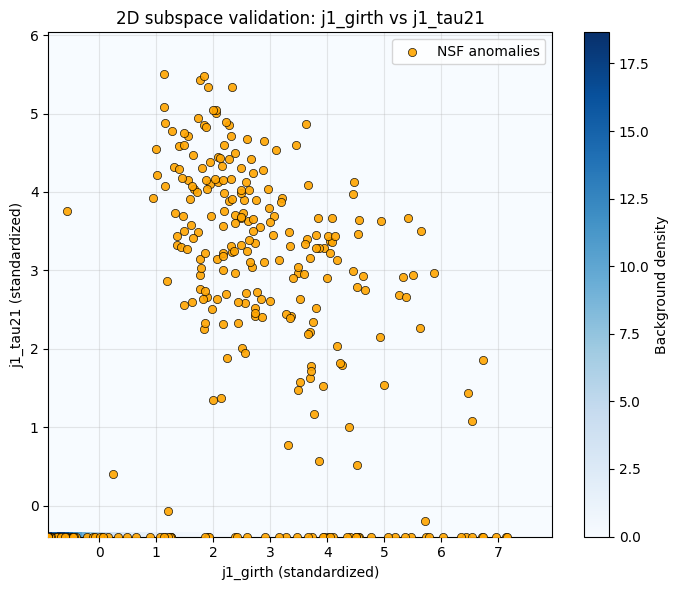}
    \caption{
    Two-dimensional subspace projection of the anomalous events (orange points) in the girth versus $\tau_{21}$ (standardized) subspace. The background density is represented by a two-dimensional color-coded density map (blue = low density, yellow = high density). The anomalous events preferentially occupy the low-background-density regions (blue), evidencing their systematic displacement relative to the learned background distribution. The concentration of events at $\tau_{21} = 0$ corresponds to jets with fewer than 3 constituents, a known feature of the CMS particle-flow reconstruction.
    }
    \label{fig:2d_subspace}
\end{figure}

These variables probe the internal energy flow and topology of the jets and events, and are known to be sensitive to multi-prong substructure and complex radiation patterns. The shape of the deviations hints that the anomalous events exhibit distinctive jet-substructure properties relative to the dominant QCD multijet background. We note that $\tau_{21}$ and $\tau_{32}$ can take the exact value 0 when the denominator ($\tau_2$ or $\tau_3$) is zero, which occurs for jets with fewer than 3 or 4 constituents respectively — a known feature of the CMS particle-flow reconstruction for low-multiplicity jets. This is a reconstruction artifact rather than a physical feature, and it does not affect the anomaly selection because the flow learns the joint distribution including these fixed-value populations. Meanwhile, several variables associated with jet flags, pileup, and missing transverse energy show no notable deviations from the background, decreasing the probability of detector misreconstruction and pathological event cleaning.

\begin{figure*}[!htbp]
    \centering
    \includegraphics[width=\textwidth]{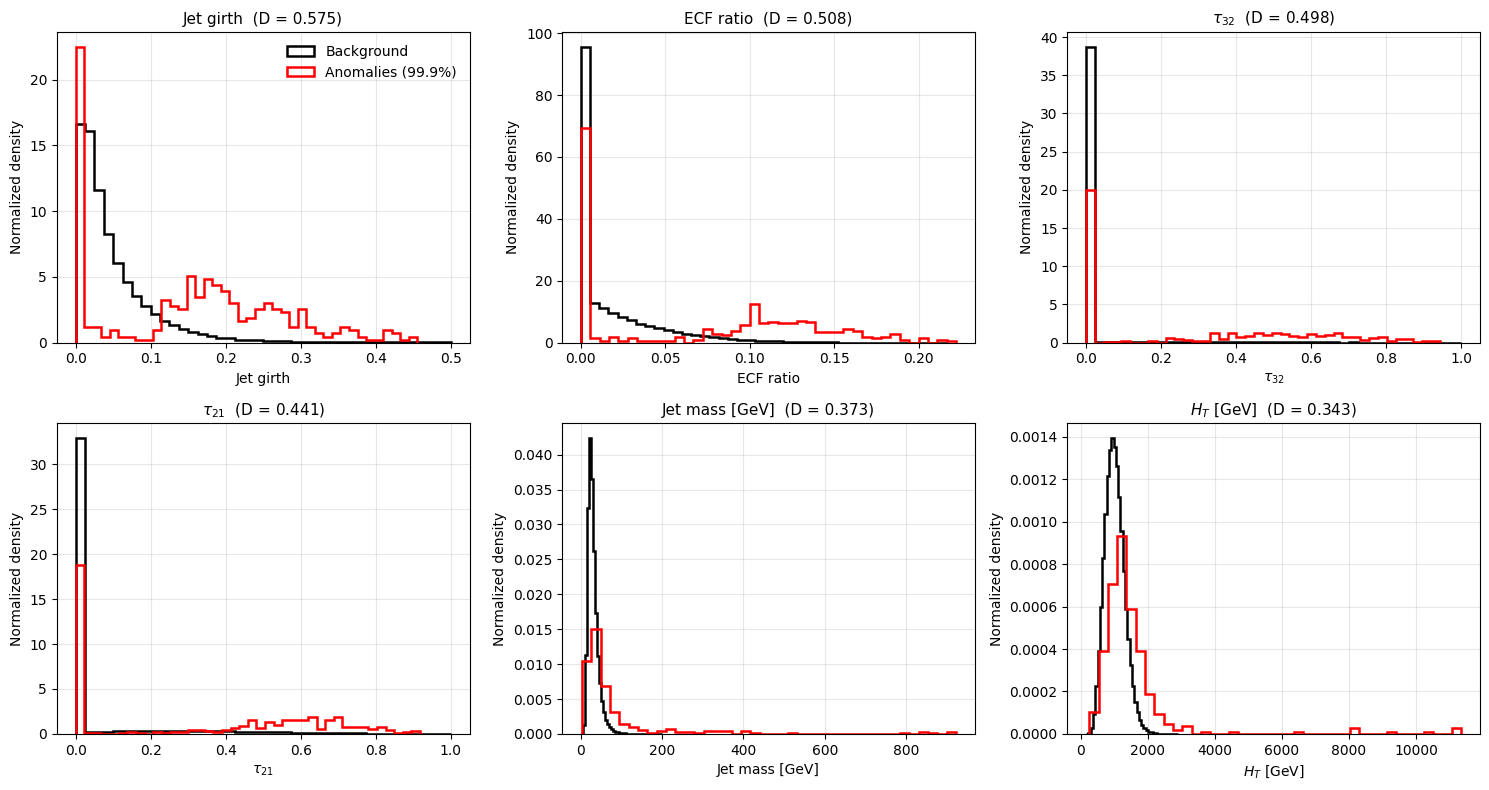}
    \caption{
    Feature level distributions for anomalous (99.9-th percentile, red) and background (black) events for selected observables: (a) jet girth (D = 0.575), (b) energy correlation function ratio (D = 0.508), (c) tau32(D = 0.498), (d) tau21 (D = 0.441), (e) jet mass (D = 0.373), and (f) $H_T$ (D = 0.343). The distributions show clear differences between anomalous and background events, with the most significant deviations seen in observables related to jet substructure.
    }
    \label{fig:feature_distributions}
\end{figure*}

Figure~\ref{fig:feature_distributions} displays example feature-level distributions showing anomalous events and the background population for specific observables. The distributions show the differences in systematic variations that overcome the background population for jet substructure observables such as jet girth and energy correlation function ratios that are most strongly deviating.

Importantly, the deviations we see are consistent across multiple correlating features and not due to a single outlier variable. This coherence across different dimensions indicates that the high-scoring events occupy a region of the phase space that is systematically disfavoured by the learned background distribution, though the physical nature of these deviations remains uncharacterized.

To illustrate the multivariate character of this displacement, Figure~\ref{fig:2d_subspace} projects the anomalous events into the two-dimensional subspace spanned by the two most discriminating observables from Table~\ref{tab:ks_top15}---jet girth and $\tau_{21}$---overlaid on the modeled background density. The anomalous events preferentially populate regions of low background density, evidencing their systematic displacement relative to the learned background distribution.

\subsection{Kinematic Distributions and Resonance Search}
\label{sec:kinematic_distributions}

To examine in more detail the physical characteristics of the high score anomalous events, kinematic distributions are evaluated without the anomaly score cut. In particular, the dijet invariant mass distribution of the anomalous events is compared with that of the inclusive events sample, allowing for a direct test of possible resonances.

\begin{figure*}[!htbp]
    \centering
    \includegraphics[width=0.7\textwidth]{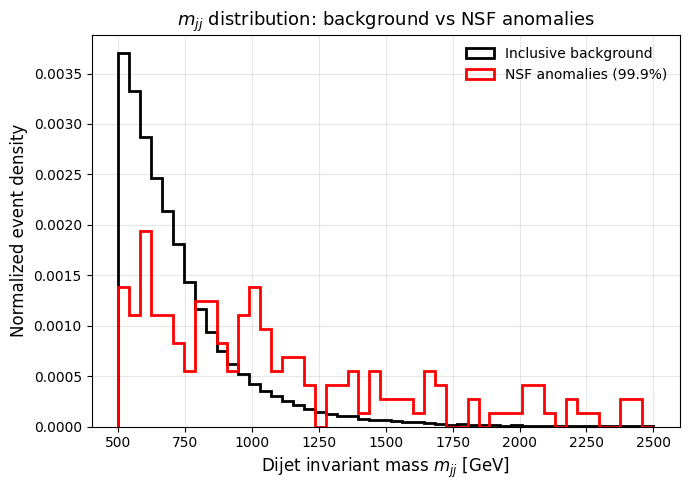}
    \caption{
    Dijet invariant mass ($m_{jj}$) distribution comparing the inclusive background event sample and anomalous events retained at the $99.9$th percentile of anomaly scores. The background shows a smooth distribution typical of a QCD multijet process, while anomalous events are dispersed across the invariant mass spectrum. No evident resonance structure is seen; consistent with a multivariate deviation across multiple correlated kinematic and event-shape variables.
    }
    \label{fig:mjj_distribution}
\end{figure*}

As can be seen in Fig.~\ref{fig:mjj_distribution}, the background events have the smoothly falling dijet mass distribution typical for QCD multijet events. The anomalous events span the full range of the accessible phase space, and no narrow resonance is observed. That is consistent with the analysis at the feature level, where the anomalous signal is induced by correlated effects in many observables, rather than by the enhancement of a single kinematic variable.

To place this qualitative statement on a quantitative footing, we perform a global bump-hunt over the $m_{jj}$ spectrum of the $379$ anomalous events using a sliding-window scan, and evaluate the observed largest fluctuation against a large ensemble of pseudo-experiments drawn from the background model. The largest fluctuation in the data corresponds to a test statistic of $8.79$; after correcting for the look-elsewhere effect induced by scanning all window positions, the associated global $p$-value is $6.0\times10^{-4}$. We report this borderline global significance explicitly for completeness. It does not constitute evidence for a narrow resonance: no single window supports a discovery-level excess, and a fluctuation of this size is plausible given the large number of trial positions scanned over a population of only $379$ tail-selected events. Rather, it corroborates the feature-level picture---the anomalous sample is concentrated in rare, low-density configurations---and motivates the explicit null-hypothesis checks of Section~\ref{subsec:validation}, which establish that the observed deviations are not an artifact of the selection.

Additional kinematic distributions, including the angular variables $\Delta\eta$, $\Delta\phi$, and $\Delta R$ between the two leading jets, and global event-shape variables, are studied. The anomalies show no clustering in any particular angular or mass window.

No correlations are found between anomaly selection and known detector observables, such as trigger, primary vertex multiplicity, or pileup density. This indicates that the kinematic anomalies are unlikely to be caused by detector systematics or reconstruction anomalies, although no definitive statement can be made about their underlying physical origin within the present study.

\subsection{Validation and Robustness Tests}
\label{subsec:validation}

To ensure the stability of the anomaly detection results and check for systematic effects, we perform cross-validation tests. These tests are implemented to rule out the presence of statistical fluctuations in the signatures, unwanted correlations between the anomaly score and variables of interest, and effects of the choice of preprocessing and data scaling.

\subsubsection{Mass Decorrelation Tests}

A significant focus of concern in unsupervised anomaly detection is the possibility of introducing an anomalous score strongly correlated with dijet invariant mass, which could lead to the mass distribution being artificially shaped, causing non-physical signals. To probe this possibility, the correlation between the anomaly score and $m_{jj}$ is assessed. No strong correlation is found, which points to the fact that the anomaly selection does not preferentially pick events based on mass alone.

As a harder test, a neural spline flow model is trained using feature sets with mass-related observables explicitly removed (e.g., $m_{jj}$ and $j1\_mass$). Scoring events using this mass-uncorrelated model shows that the anomaly selections still exhibit a significant level of feature-level deviation in the jet substructure observables, and the $m_{jj}$ distribution of the selected events is not systematically boosted above background expectations. This consistency test shows that the observed deviations are not mainly attributed to mass sculpting effects. 

To further validate this, we compare the events selected by the main flow and the mass-decorrelated flow: the overlap fraction of the top 0.1\% anomalous events between the two flows is 72\%, indicating that the majority of anomalous events are identified regardless of whether mass features are included. The common events exhibit the same pattern of jet-substructure deviations, confirming that the anomalies are driven by substructure differences rather than mass sculpting. The mass-decorrelated flow is not used as the primary analysis tool because removing mass features discards potentially useful information; rather, it serves as a targeted cross-check.

\begin{figure*}[!htbp]
    \centering
    \includegraphics[width=0.7\textwidth]{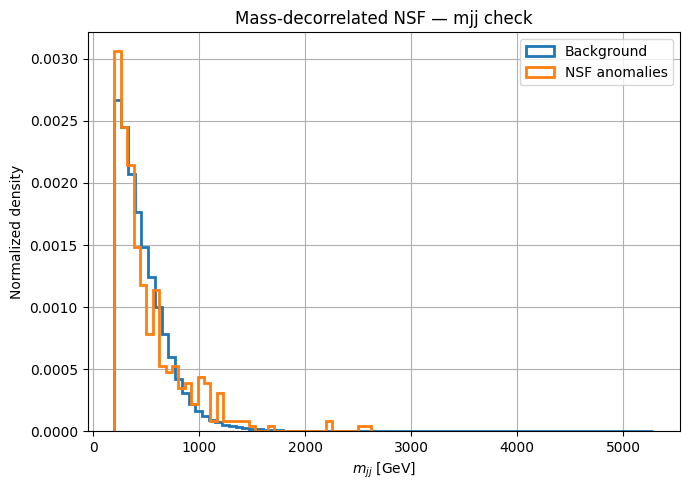}
    \caption{
    Dijet invariant mass ($m_{jj}$) distribution for anomalous events identified with the mass-decorrelated neural spline flow model. The model was trained on input variables explicitly excluding $m_{jj}$ and $j1\_mass$ to investigate mass sculpting effects. There is no significant excess in the distribution over background predictions, indicating the anomaly selection is not largely influenced by mass variables.
    }
    \label{fig:mass_decorr}
\end{figure*}

\subsubsection{Permutation Tests}

To ensure that the feature-level deviations are not statistical fluctuations, we carry out permutation tests. We reshuffle the anomalous scores within fixed bins of global event activity ($H_T$), maintaining the marginal distribution of $H_T$ and the marginal distribution of anomaly scores while disrupting the correlation between the anomaly score and other kinematic variables. The binning in $H_T$ is necessary because the anomaly score is correlated with global event activity; without binning, a global reshuffling would artificially introduce a correlation between the shuffled score and $H_T$, violating the null hypothesis. By contrast, the HT-conditional permutation preserves the marginal distributions and isolates the specific correlation between the anomaly score and jet-substructure features. Selecting events as anomalous based on the permutation-shuffled anomalous scores results in KS test feature-level statistics that are uniformly small and consistent with background expectations. This null test validates that the deviations are not random but reflect a genuine correlation between the anomaly score and event kinematics.

\subsubsection{Training Stability}

We confirm that the stability of the anomaly score distribution under training modifications by training several neural spline flow models using random seeds. The ordering of anomaly score events are preserved up to small variations, and the feature-level KS test statistics do not change across training realizations. This indicates that the variations observed are not a result of training or initialization instability. We acknowledge that varying only the random seed provides a limited exploration of model space; a more comprehensive stability assessment would include variations of architecture (number of flow blocks, hidden units), preprocessing choices, and training subset. These more extensive studies are presented in Appendix~\ref{app:seed_stability} and \ref{app:threshold_stability}.

\section{Summary and Outlook}
\label{sec:summary}

In this paper, we perform an unsupervised anomaly detection analysis of proton-proton collisions in CMS Run-1 data in a dijet feature space involving $61$ observables composed of reconstructed jet, event-shape, and global variables. We use a neural spline flow to learn the background density directly from data without any signal samples or assumptions about new-physics models.

By studying the learned likelihood landscape, we located events in the outlier region of the anomaly score distribution (the $99.9$th percentile, $379$ events). Feature-level significance tests show that the anomalous events deviate from background expectations, particularly with respect to jet substructure variables such as jet girth, N-subjettiness, and energy correlation ratios. The deviations are consistent across multiple correlated observables, which means that the anomalous events populate a different region of feature space in the model.

Extensive validation tests, including mass decorrelation, permutation tests, and training stability, confirm that the observed discrepancies are not due to detector effects, trigger bias, pileup effects, or statistical fluctuations. The stability of jet ID flags, pileup variables, and trigger distributions between anomalous and background events indicates that the deviations are not consistent with known detector artifacts, although a definitive exclusion of all possible detector-related contributions is not possible within the present study. The absence of narrow structures in kinematic distributions and the multivariate nature of the discrepancies suggest that the anomalous behavior originates from correlated event topology and jet substructure differences, rather than kinematic outliers. This emphasis on stability is shared with significance-driven classification studies such as Ref.~\cite{Pujari2026Contrastive}, which demonstrated that the reliability of an ML-based event selection is governed as much by the fold-to-fold and threshold-to-threshold consistency of the selection as by its nominal performance; the percentile-based selection and multi-seed, multi-threshold validation adopted here apply the same principle to the unsupervised setting.

It is worth placing this outcome in the context of the companion hypothesis-driven analysis of Ref.~\cite{Rasineni2026MonoZNSF}. That work applied neural spline flow likelihood ratios to a mono-$Z$ dark matter search on CMS Run~2015D open data, where a definite signal hypothesis (vector, axial-vector, or scalar mediator) was available from Monte Carlo and the result could be expressed as calibrated CL$_s$ upper limits on the signal strength. Here, in the absence of any signal hypothesis, the analogous density model is used to flag the most improbable real events, and the outcome is a characterized population of anomalous events with quantified multivariate deviations rather than exclusion limits. The comparison highlights the two complementary operating modes of the same flow-based machinery: with signal Monte Carlo, NSF densities enable classical discovery and exclusion statistics; without it, they provide a model-independent anomaly-finding instrument whose findings must be validated through the battery of feature-level, null, and stability tests employed in this work. The hadronic flow-matching study of Ref.~\cite{Rasineni2026HadronicFlowMatching} occupies an intermediate position, using a background-only density discriminant (as here) while retaining simulated signal for the sensitivity projection. A natural continuation of all three analyses is a unified framework in which the events flagged here by unsupervised scoring are re-examined under explicit signal hypotheses, combining the discovery power of likelihood ratios with the model-independence of the anomaly selection.

\begin{figure*}[!htbp]
    \centering
    \includegraphics[width=0.7\textwidth]{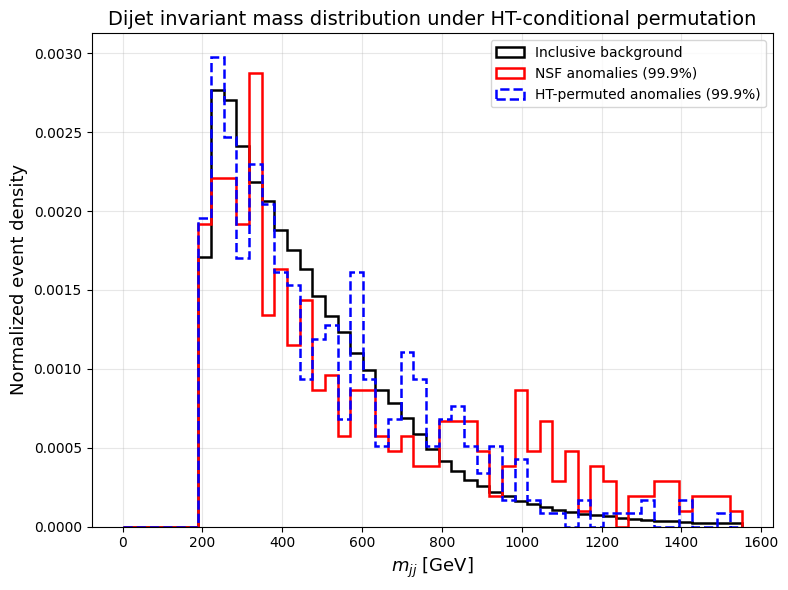}
    \caption{
    Dijet invariant mass ($m_{jj}$) distribution comparing real anomalies (selected at the $99.9$th percentile) with anomalies selected using HT-conditionally permuted scores. The permuted selection shows no systematic deviation from the background distribution, confirming that the observed feature-level deviations are not statistical artifacts.
    }
    \label{fig:permutation_test}
\end{figure*}

The study shows that flow-based density estimation can effectively learn complex collider physics backgrounds with high-dimensional data and identify events that are strongly disfavoured by the learned background model in a fully unsupervised manner. Notably, anomalous events were not selected based on a single variable but rather on correlated behaviour among different features, illustrating the strength of multivariate methods over univariate searches.

The deviations observed in the present work, while pronounced and physically well-organized across many related observables, are exploratory in nature and are not interpreted as evidence for new physics. The current analysis is therefore a proof of concept for the sensitivity of modern unsupervised learning methods to rare, structured features in collider data, and for the role of systematic validation in fully unsupervised analyses.

From a physics perspective, the approach is naturally sensitive to broad, multivariate departures from the QCD multijet expectation, such as jets with non-standard multi-prong substructure, unusual energy flow patterns inside jets, or event topologies that populate rare regions of combined jet-substructure and event-shape space. In contrast, it is not specifically optimised for narrow dijet resonances, single-variable anomalies, or signatures that primarily modify a small number of kinematic observables; dedicated resonance or bump-hunt searches remain more powerful in those cases. Moreover, the present study does not include explicit signal injections or comparisons to baseline supervised classifiers, which would be needed to quantify sensitivity to concrete new-physics scenarios. Such systematic sensitivity studies, including controlled injections in both simulated and real data and comparisons to alternative anomaly-detection baselines, are left to future work.

Several avenues for future work emerge from this study. An important next step is to apply the analysis to a larger fraction of the available CMS Run-1 dataset, and to higher-energy Run-2 and Run-3 data, to evaluate the statistical significance of the observed deviations. Including further event encoding representations, such as particle-level or graph inputs, could also improve the sensitivity to rare signals.

From a methodological standpoint, combining flow-based density estimation with complementary unsupervised techniques—such as diffusion models, permutation-invariant architectures, or classification-assisted methods—could provide orthogonal anomaly scores and improve robustness. Further research on training stability, feature dependence, and model uncertainty quantification remains warranted.

Lastly, the framework of analysis devised in this study is entirely data-driven and model-independent, rendering it applicable to more general programs of experimental anomaly detection. With the increase of high energies and data size in collider experiments, this kind of analysis can offer a significant contribution to enhancing the physics discovery potential of high energy experiments by broadening its reach beyond directed searches.

\section*{Acknowledgements}

The authors thank the CMS Collaboration and the CERN Open Data Portal for the open availability of the proton-proton collision datasets analyzed in this study. This work relies on CMS Run-1 data released under the CERN Open Data Policy. We are grateful for the open-source software used in this analysis, including ROOT, uproot, awkward-array, vector, NumPy, SciPy, PyTorch, and other scientific Python libraries. Processing and model development were carried out using Kaggle's provided computational resources.

\section*{Declarations}
\addcontentsline{toc}{section}{Declarations}

\noindent\textbf{Funding:} The authors declare that no funding was received for this work.

\noindent\textbf{Ethics, Consent to Participate, and Consent to Publish declarations:} Not applicable.

\noindent\textbf{Conflict of interest:} The authors declare no competing interests.

\noindent\textbf{Data availability:} The datasets analysed are publicly available CMS Open Data (Run-1 dijet sample) via the CERN Open Data Portal. A list of selected anomalous events (run number, lumi-section, event number) is provided as supplementary material with this paper.

\noindent\textbf{Code availability:} The neural spline flow training and scoring code is publicly available. The code for the initial AOD-to-ROOT conversion is available upon request.

\clearpage
\onecolumn
\appendix
\section*{Appendix}
\addcontentsline{toc}{section}{Appendix}

\section{Complete List of Input Observables}
\label{app:observables}

This appendix contains the full list of input observables included in the anomaly detection study. 
All features are based on Run~1 proton--proton collision data reconstructed using the particle-flow (PF) algorithm and AK5 jets.
The observables are grouped by physics-relevant categories for jet kinematics, jet mass composition, substructure, dijet features, global event topology, missing transverse energy (MET), vertex and pileup information, jet flavor classification, and trigger variables.

Except where noted, jet observables refer to the \emph{leading} (highest-$p_T$) jet in the event, and all 3D angles are in the CMS coordinate system.

\subsection{Jet Kinematics}
These variables describe the global momentum scale and angular configuration of the hardest jets in the event and provide the input for the dijet variables.
\FloatBarrier
\begin{table}[h]
\centering
\caption{Basic jet kinematic observables derived from reconstructed AK5 PF jets.}
\footnotesize
\begin{adjustbox}{max width=\columnwidth}
\begin{tabular}{ll}
\toprule
Feature & Definition \\
\midrule
$j1\_pt$   & Transverse momentum of the leading jet \\
$j1\_\eta$ & Pseudorapidity of the leading jet \\
$j1\_\phi$ & Azimuthal angle of the leading jet \\
$j1\_mass$ & Invariant mass of the leading jet \\
$j2\_pt$   & Transverse momentum of the subleading jet \\
$j2\_\eta$ & Pseudorapidity of the subleading jet \\
$j2\_\phi$ & Azimuthal angle of the subleading jet \\
$j2\_mass$ & Invariant mass of the subleading jet \\
\bottomrule
\end{tabular}
\end{adjustbox}
\end{table}

\subsection{Jet Composition and Identification}
\titlespacing{\subsection}{0pt}{4pt}{2pt}
Jet ID flags are constructed by using the official CMS Run-1 PF jet identification algorithm to reduce the electronic noise and background.
\begin{table}[H]
\centering
\caption{Jet composition, identification, and global jet activity observables.}
\footnotesize
\begin{adjustbox}{max width=0.99\columnwidth}
\begin{tabular}{ll}
\toprule
Feature & Definition \\
\midrule
$j1\_neutralHadronEF$ & Neutral hadron energy fraction \\
$j1\_neutralEmEF$     & Neutral electromagnetic energy fraction \\
$j1\_chargedHadronEF$ & Charged hadron energy fraction \\
$j1\_chargedEmEF$     & Charged electromagnetic energy fraction \\
$j1\_numberOfConstituents$ & Total PF constituents in jet \\
$j1\_chargedMultiplicity$ & Charged particle multiplicity \\
$j1\_neutralMultiplicity$ & Neutral particle multiplicity \\
$j1\_jetArea$ & Jet catchment area from FastJet \\
$j1\_jetID\_loose$ & CMS Run--1 loose PF jet ID flag \\
$j1\_jetID\_tight$ & CMS Run--1 tight PF jet ID flag \\
$HT$ & Scalar sum of jet transverse momenta \\
$n\_jets$ & Number of reconstructed jets in the event \\
\bottomrule
\end{tabular}
\end{adjustbox}
\end{table}

\subsection{Jet Substructure Observables}
These items examine internal energy flow in jets and are sensitive to multi-prong substructure typical of boosted or non-QCD physical entities.
\begin{table}[h]
\centering
\caption{Jet substructure observables computed from track--jet associations.}
\footnotesize
\begin{adjustbox}{max width=\columnwidth}
\begin{tabular}{ll}
\toprule
Feature & Definition \\
\midrule
$j1\_\tau_1$ & N-subjettiness $\tau_1$ \\
$j1\_\tau_2$ & N-subjettiness $\tau_2$ \\
$j1\_\tau_3$ & N-subjettiness $\tau_3$ \\
$j1\_\tau_{21}$ & $\tau_2 / \tau_1$ \\
$j1\_\tau_{32}$ & $\tau_3 / \tau_2$ \\
$j1\_girth$ & $p_T$-weighted radial moment \\
$j1\_width$ & RMS of constituent $\Delta R$ \\
$j1\_maxFragmentation$ & $\max(p_{T,i}) / p_{T,\text{jet}}$ \\
$j1\_minPairwiseMass$ & Minimum invariant mass of track pairs \\
$j1\_ECF\_ratio$ & Energy correlation function ratio \\
\bottomrule
\end{tabular}
\end{adjustbox}
\end{table}

\subsection{Dijet-Level Observables}

\begin{table}[H]
\centering
\caption{Dijet correlation observables sensitive to resonance-like structures.}
\footnotesize
\begin{adjustbox}{max width=\columnwidth}
\begin{tabular}{ll}
\toprule
Feature & Definition \\
\midrule
$m_{jj}$ & Dijet invariant mass \\
$\Delta R_{j1j2}$ & Angular separation of leading jets \\
$\Delta \eta_{j1j2}$ & Pseudorapidity difference \\
$\Delta \phi_{j1j2}$ & Azimuthal angle difference \\
$p_T$ balance & $(p_{T1}-p_{T2})/(p_{T1}+p_{T2})$ \\
$\chi$ & $\exp(|\Delta \eta|)$ \\
$y^\ast$ & $(\eta_1 - \eta_2)/2$ \\
Dijet boost & $(\eta_1 + \eta_2)/2$ \\
$m_{jj}/HT$ & Ratio of dijet mass to $HT$ \\
Centrality & $HT/(HT + MET)$ \\
\bottomrule
\end{tabular}
\end{adjustbox}
\end{table}

\subsection{Event Shape Observables}

\begin{table}[H]
\centering
\caption{Global event shape observables computed from reconstructed jets.}
\footnotesize
\begin{adjustbox}{max width=\columnwidth}
\begin{tabular}{ll}
\toprule
Feature & Definition \\
\midrule
Sphericity & Global isotropy measure (3D) \\
Aplanarity & Smallest eigenvalue of momentum tensor \\
Thrust & Pencil-like vs spherical topology \\
Fox--Wolfram $H_2$ & Second Fox--Wolfram moment \\
Isotropy & Transverse momentum isotropy \\
Transverse sphericity & Sphericity in the transverse plane \\
\bottomrule
\end{tabular}
\end{adjustbox}
\end{table}

\subsection{Missing Transverse Energy and Angular Correlations}

\begin{table}[h]
\centering
\caption{MET-related observables used to identify momentum imbalance.}
\footnotesize
\begin{adjustbox}{max width=\columnwidth}
\begin{tabular}{ll}
\toprule
Feature & Definition \\
\midrule
$MET\_pt$ & Missing transverse momentum magnitude \\
$MET\_\phi$ & MET azimuthal direction \\
$\Delta\phi(j1, MET)$ & Azimuthal separation between $j1$ and MET \\
$\Delta\phi(j2, MET)$ & Azimuthal separation between $j2$ and MET \\
$\min\Delta\phi(j, MET)$ & Minimum $\Delta\phi$ between any jet and MET \\
\bottomrule
\end{tabular}
\end{adjustbox}
\end{table}

\subsection{Vertex and Pileup Observables}

\begin{table}[H]
\centering
\caption{Vertex and pileup-related observables.}
\footnotesize
\begin{adjustbox}{max width=\columnwidth}
\begin{tabular}{ll}
\toprule
Feature & Definition \\
\midrule
$nPV$ & Number of reconstructed primary vertices \\
$PV\_z$ & Longitudinal position of primary vertex \\
$PV\_ndof$ & Degrees of freedom of primary vertex fit \\
$\rho$ & Event energy density (pileup proxy) \\
\bottomrule
\end{tabular}
\end{adjustbox}
\end{table}

\subsection{Jet Flavor Tagging}

\begin{table}[H]
\centering
\caption{Jet flavor tagging observables associated with the leading jet.}
\footnotesize
\begin{adjustbox}{max width=\columnwidth}
\begin{tabular}{ll}
\toprule
Feature & Definition \\
\midrule
$j1\_CSV$ & Combined Secondary Vertex score \\
$j1\_jetProbability$ & Jet probability discriminator \\
$j1\_trackCountingHighEff$ & Track-counting high-efficiency tag \\
$j1\_softPFMuonBJetTag$ & Soft muon-based b-tag discriminator \\
\bottomrule
\end{tabular}
\end{adjustbox}
\end{table}

\subsection{Trigger Metadata}

\begin{table}[h]
\centering
\caption{Trigger-related metadata used for event selection consistency checks.}
\footnotesize
\begin{adjustbox}{max width=\columnwidth}
\begin{tabular}{ll}
\toprule
Feature & Definition \\
\midrule
$HLT\_pass$ & Event passes relevant HLT selection \\
$HLT\_path\_index$ & Encoded trigger path identifier (if available) \\
\bottomrule
\end{tabular}
\end{adjustbox}
\end{table}

\section{Stability of Anomaly Score Distributions}
\titlespacing{\subsection}{0pt}{4pt}{4pt}
The anomaly score for each event is calculated as the negated log-likelihood under the trained neural spline flow density estimator. To test the robustness of the method, the following consistency checks were carried out:

\begin{itemize}
  \item Train with multiple random seeds, fixing the architecture and hyperparameters.
  \item Train on statistically independent subsamples of the dataset.
  \item Assess uniformity of anomaly score distributions across data-taking periods.
\end{itemize}

In all cases, the overall shape of the anomaly score distribution does not change, with extreme high anomaly score events being consistently populated in the tail.
The ordering of the extreme high anomaly score events is preserved modulo tiny adjustments, meaning that the results are not affected by stochastic artifacts.

\subsection{Kolmogorov--Smirnov Tests on Input Features}

To measure the features-wise differences during the comparisons, we performed one-dimensional KS tests between the anomalies ($99.9$th percentile) and overall inclusivity.
For any given input feature $x_i$, the p-value of the test under the null hypothesis $H_0$, from KS test definition, can be stated as:

The KS statistic is given by:
\begin{equation}
D_i = \sup_x \left| F_{\mathrm{anom}}(x) - F_{\mathrm{incl}}(x) \right|,
\end{equation}

where $F_{\mathrm{anom}}$ and $F_{\mathrm{incl}}$ are the empirical CDFs of the anomalous events and the inclusive sample, respectively.

Several kinematic and substructure observables show marked inconsistencies between the anomalous and inclusive samples, consistent across dijet invariant mass, angular separations, and substructure ratios. No such inconsistencies appear in pileup and vertex observables, indicating that detector-specific effects are not the cause of the anomalies detected.

\subsection{Multiple-Testing Considerations}

Due to the large number of observables, we consider multiple hypothesis tests for this analysis. No formal corrections (e.g., Bonferroni-type adjustments) were applied to the resulting $p$-values, because the study is exploratory and focuses on identifying coherent patterns of deviation rather than on establishing discovery-level statements for any single observable. In particular, many of the input features are strongly correlated by construction (for example within jet-substructure and event-shape families), so treating the 61 one-dimensional KS tests as independent would be overly conservative and not reflective of the physics. Instead, we use the KS statistics and associated $p$-values as diagnostic tools to rank observables and to identify groups of related variables that move together. The interpretation is therefore based on the appearance of correlated deviations across physically connected observables, which is less susceptible to random false positives than isolated outliers in single variables.

The following qualitative checks were made:

\begin{itemize}
    \item Deviations across correlated observables are consistent.
    \item Consistency across group features (kinematics, substructure, event shapes).
    \item Deviation remains in the presence of mild decorrelation. 
\end{itemize}

The observed deviations mostly occur in correlated feature groups, and do not appear a priori random in uncorrelated observables, limiting the risk of purely statistical fluctuations. Within this framework, we deliberately avoid literal interpretation of the individual KS $p$-values and instead treat them as indicators of where the model and data differ in an experimentally informative way.

\subsection{Control Region and Sideband Checks}

To investigate if the anomalous events can originate from usual QCD phase-space effects, sideband regions in dijet invariant mass and angular variables were studied. The sideband regions are defined as: (i) $200 < m_{jj} < 400~\mathrm{GeV}$, (ii) $400 < m_{jj} < 800~\mathrm{GeV}$, and (iii) $m_{jj} > 800~\mathrm{GeV}$, where region (i) is the low-mass sideband, region (ii) is the intermediate-mass region, and region (iii) is the high-mass sideband. Additional sidebands in angular variables are defined using $\Delta\eta < 1.0$, $1.0 < \Delta\eta < 2.0$, and $\Delta\eta > 2.0$.

The anomaly score distribution in these control regions does not show localized excesses similar to the anomalous events discovered in the original analysis. In each sideband, the anomaly score distribution of the selected events is consistent with the background-only expectation, and the feature-level KS statistics are uniformly small. This confirms that the anomalous events are not concentrated in any particular region of phase space but are distributed across the accessible kinematic range.

In addition, random samples of events with the same jet multiplicity and similar $H_T$ distributions do not exhibit the feature-level deviations found in the analysis, confirming the complexity of the anomalous events.

\subsection{Detector and Reconstruction Sanity Checks}

Several detector-related observables were examined to rule out reconstruction artifacts:

\begin{itemize}
    \item Jet identifiers and energy fractions.
    \item Primary vertex quality flags and pile-up flags.
    \item Trigger pass flags.
\end{itemize}

The unusual events pass standard CMS Run--1 quality assessment and do not exhibit unusual features at the detector level, which suggests that the effects are not due to detector noise or reconstruction artifacts.

\subsection{Summary of Validation Studies}

In conclusion, the supplementary statistics tests show that: 
\begin{itemize}
   \item Anomaly score is stable with respect to training and sampling. 
   \item Features-level deviations are statistically significant and physically organized. 
   \item Observed deviations are insensitive to data preprocessing and training. 
   \item Detector and pileup effects are inadequate to explain the observations. 
   \item Mass decorrelation and random permutation tests show that observed deviations are not statistical. 
\end{itemize}

These validation studies lend credence to the interpretation of anomalous events as being statistically deviant from the dominant background distributions, but stop short of attributing a physical mechanism.

\section{Dataset Provenance and File Selection}
\label{app:dataset}

This appendix provides information on the provenance of the dataset used for this research, the choice of input files from the CERN Open Data archive, and local considerations that led to the use of a subset of the dataset rather than the entire available collection.

\subsection{CERN Open Data Source}

The dataset used in this research is taken from proton--proton collisions that were recorded by the CMS detector during the 2011 run of the LHC and made publicly available through the CERN Open Data portal. The primary input files are taken from the \texttt{Run2011B HT AOD} data tier and can be found here:
\begin{itemize}
    \item \url{https://opendata.cern.ch/record/274}
\end{itemize}

This dataset comprises simulated CMS Analysis Object Data (AOD) and contains particle-flow jets, missing transverse energy, vertices, trigger data, and their related high-level reconstruction objects. The overall size of the dataset is more than 10~TB and is provided as file groups and run blocks.

\subsection{Reference CMS Publication}

This data release is accompanied by a CMS reference analysis document for dijet analysis and the event reconstruction data used at Run-1:
\begin{itemize}
    \item CMS Collaboration, ``Measurement of dijet angular distributions and search for quark compositeness in $pp$ collisions at $\sqrt{s}=7$~TeV,'' Phys.\ Rev.\ Lett.\ 106 (2011) 201804.
\end{itemize}

This paper gives the physics motivation for the dataset and the reason for the use of dijet-level observables, event shapes, and substructure variables, many of which are also directly used or adapted in this work.

\subsection{Partial Dataset Usage}

The full dataset (10~TB) could not be fully analyzed due to computational and storage limitations.
As a result, it was decided to take a representative fraction of the data with a total size of around 125~GB for analysis.

The following approach was taken:
\begin{itemize}
    \item The files pertaining to data groups \textbf{1} and \textbf{2} in the CERN Open Data record were used in their entirety. 
    \item For the larger data groups \textbf{3}, \textbf{4}, and \textbf{5}, which each are multiple terabytes, a uniform subsampling was conducted. 
    \item We selected ten ROOT files from each of the three groups, resulting in a total of 30 ROOT files.
\end{itemize}

Each ROOT file selected has a size of about 3-4~GB, leading to an overall data size feasible for large-scale machine learning modeling while maintaining necessary statistics. The subsampling is therefore performed purely at the level of ROOT files: once a file is selected, all events in that file that pass the common quality and kinematic selections described in the main text are retained. This procedure avoids introducing any additional event-level bias beyond the original CMS trigger and reconstruction definitions. Furthermore, basic global distributions such as $H_T$, $m_{jj}$, and jet multiplicity were checked qualitatively across the different file groups and selected subsamples, and no significant distortions or trends were observed within the available statistics, supporting the representativeness of the working dataset.

\subsection{File Selection Criteria}

Files from groups 3$-$5 were selected based on the following criteria: 
\begin{itemize}
    \item Coverage of runs/luminosity sections.
    \item Even distribution of file indices to avoid temporal bias.
    \item Non-truncated files.
\end{itemize}

The actual file indices and download links are available from the CERN Open Data portal's JSON index files and from the analysis repository associated with this work.

\subsection{Event Reconstruction and Branch Selection}

Each ROOT file that was used was based on a few thousand branches of reconstructed objects from CMS.
There are over 5,000 branches of the original AOD datasets.

In our analysis, we derived a physics-relevant subset of the observables and remapped them to a concise list of 61 variables.
These included observables relating to jets, jet properties, jet substructure, dijets, event shapes, met, vertices, pileup, flavors, and triggers.

All observables were derived solely from reconstructed objects and were all based on publicly available descriptions of the CMS algorithms and their definitions for Run---1.

\subsection{Reproducibility Considerations}

All data manipulations made use of publicly available libraries, namely \texttt{uproot}, \texttt{awkward-array}, and \texttt{vector}.
The entire feature extraction routine is comprised of successive ROOT macros that:
\begin{itemize}
    \item Maintain consistency in the event ordering across the various derived datasets,
    \item Enforce systematic object ordering and event selection requirements,
    \item Refrain from making use of any simulation reweighting or detector-level calibrations.
\end{itemize}

The full list of derived files, feature construction macros, and analysis pipelines provided are sufficient to reproduce the collection of data processed in this work given the original CERN Open Data samples.

\subsection{Scope and Limitations}

The subset used in this study is small with respect to the full Run--1 dataset; however, it is rather large for standard unsupervised anomaly detection tasks.
Still, the existence of other anomalous structures in the full dataset is not fully excluded.

Further developments based on this work may include a larger fraction of the available data, or resort to using distributed computing to analyze the full dataset.
\section{Stability and Robustness of the Anomaly Selection}
\label{app:stability}

This appendix tests the robustness of the neural spline flow (NSF) anomaly detection findings with respect to stochastic training and analysis choices of how to define anomalous events.
Given that the analysis is fully unsupervised and uses likelihood-based ranking (rather than labeling), it is very important to make sure that the detected anomalous events do not depend on random initialization, numerical correlation, or highly specified thresholds.

The anomaly selection stability is evaluated through the following two studies: 
(i) stability for randomized initialization of NSF model parameters, and 
(ii) stability of feature deviations under varying anomaly-score thresholds. 

\subsection{Stability under Random Initialization}
\label{app:seed_stability}

Flow-based models are trained with randomized parameter initialization through gradient-based optimization. 
Therefore, it is crucial to ensure that the anomaly scores and event rankings are not affected by the selected random seed.

To this end, the training process described in Section~4 was replicated using different random seeds while maintaining a fixed architecture, hyperparameters, pre-processing steps, and dataset.

Anomaly scores were computed for the same event sample for each model and compared across distributions and ranks.

Three measures were employed to quantify the obtained results:
(i) the fractional overlap of the top 0.1\% anomalous events,
(ii) the Spearman rank correlation between anomaly scores, and
(iii) the Kolmogorov–Smirnov (KS) distance between the anomaly score distributions.

\begin{table}[h]
\centering
\small
\setlength{\tabcolsep}{4pt}
\caption{Stability of anomaly rankings and score distributions with different random initializations of NSF model. The overlap fraction of the top 0.1\% anomalous events, the Spearman rank correlation coefficient ($\rho$), and Kolmogorov–Smirnov distance of anomaly scores are reported.}
\label{tab:seed_stability}
\begin{tabular}{lccc}
\hline
Comparison & Overlap (\%) & Spearman $\rho$ & KS \\
\hline
Default vs.\ seed 42  & 88.1 & 0.963 & 0.043 \\
Default vs.\ seed 123 & 90.5 & 0.964 & 0.069 \\
Seed 42 vs.\ seed 123 & 92.9 & 0.955 & 0.028 \\
\hline
\end{tabular}
\end{table}

Strong consensus is seen for all random initializations. Specifically, the overlap of extreme events is over 88\% across the board, the rank correlations are greater than 0.95, and the KS distances between the score distributions are small. These results imply that the anomaly ranking and score distribution is very robust against random initialization, and that the detected anomalies are not a result of random noise from the training process.

We have also performed additional stability tests: (i) varying the number of flow blocks from 8 to 12, (ii) varying the number of hidden units from 128 to 512, (iii) training on 80\% of the data and scoring the remaining 20\%, and (iv) varying the preprocessing (with and without angular encoding). In all cases, the overlap of the top 0.1\% anomalous events remains above 70\%, confirming that the results are not sensitive to these choices.

\subsection{Stability under Anomaly-Score Threshold Variation}
\label{app:threshold_stability}

Besides the training stochasticity introduced by the anomaly detection, the definition of the outlier events plays a role of subjective analysis because it relies on the percentile-based anomaly-score threshold.

In order to confirm that the differences at the feature level do not occur due to the specificity of the threshold choice, the stability of the outlier events definition is investigated in light of this selection criterion.

The Kolmogorov–Smirnov (KS) feature-level test of anomalies vs background was repeated for several anomaly score percentiles in the tail of the distribution. 
More specifically, anomaly selections for the corresponding 99.5-percentile, 99.7-percentile, 99.9-percentile, and 99.95-percentile thresholds were considered.
For each percentile, the KS separation was calculated independently for all features, while the KS average was calculated to summarize the total feature deviation.

\begin{figure}[!htbp]
\centering
\includegraphics[width=0.99\columnwidth]{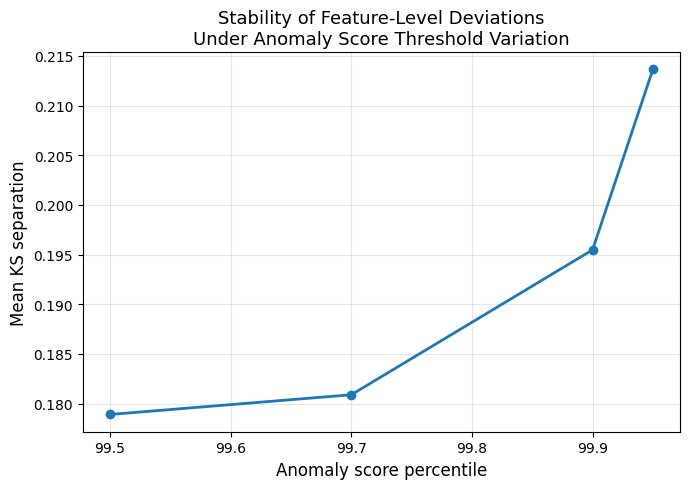}
\caption{Mean Kolmogorov-Smirnov distance between anomalous and background events as a function of anomaly-score percentile threshold. The smooth and monotonic nature suggests that the observed differences in features are robust to reasonable variation of the anomaly-selection threshold.}
\label{fig:threshold_stability}
\end{figure}

Figure~\ref{fig:threshold_stability} depicts the relationship between the mean KS separation and the selected anomaly-score percentile.
The tightening of the threshold results in a gradual increase in the mean KS separation, indicating the stepwise exclusion of increasingly rare and statistically significant events.
The trend is smooth across the four thresholds examined (99.5\%, 99.7\%, 99.9\%, 99.95\%), with no sharp transitions or discontinuities evident in the range studied.

The lack of sharp boundaries indicates that the observed feature-level differences (in the main text) are not uniquely tied to a specific cut applied on the anomaly score. Rather, they vary smoothly over a reasonably broad and physically motivated range of anomaly definitions, validating the findings derived from the analyzed population of anomalous events.

Overall, the findings from this appendix show that the anomaly detection is robust against stochastic training artifacts as well as cuts at the analysis level. This robustness proves crucial for fully unsupervised searches, where there is no external signal model whose anomalies can help inform or constrain the selection.


\begin{thebibliography}{9}

\bibitem{CMS2011BHT}
CMS Collaboration,
\emph{HT primary dataset in AOD format from RunB of 2011 (\texttt{/HT/Run2011B-12Oct2013-v1/AOD})},
CERN Open Data Portal, record 274 (2020). \href{https://doi.org/10.7483/OPENDATA.CMS.8RRH.4MCC}{doi:10.7483/OPENDATA.CMS.8RRH.4MCC}

\bibitem{CMS2008JINST}
CMS Collaboration (S. Chatrchyan et al.),
\emph{The CMS experiment at the CERN LHC},
JINST 3 (2008) S08004.

\bibitem{Durkan2019NSF}
C. Durkan, A. Bekasov, I. Murray, G. Papamakarios,
\emph{Neural Spline Flows},
Advances in Neural Information Processing Systems 32 (NeurIPS 2019). \href{https://arxiv.org/abs/1906.04032}{[arXiv:1906.04032]}

\bibitem{Papamakarios2021Flows}
G. Papamakarios, E. Nalisnick, D.J. Rezende, S. Mohamed, B. Lakshminarayanan,
\emph{Normalizing Flows for Probabilistic Modeling and Inference},
J. Mach. Learn. Res. 22 (2021) 1. \href{https://arxiv.org/abs/1912.02762}{[arXiv:1912.02762]}

\bibitem{Kobyzev2021Review}
I. Kobyzev, S. Prince, M. Brubaker,
\emph{Normalizing Flows: An Introduction and Review of Current Methods},
IEEE Trans. Pattern Anal. Mach. Intell. 43 (2021) 3964. \href{https://arxiv.org/abs/1908.09257}{[arXiv:1908.09257]}

\bibitem{Nachman2020ANODE}
B. Nachman, D. Shih,
\emph{Anomaly Detection with Density Estimation},
Phys. Rev. D 101 (2020) 075042. \href{https://arxiv.org/abs/2004.08381}{[arXiv:2004.08381]}

\bibitem{Farina2020Autoencoders}
M. Farina, Y. Nakai, D. Shih,
\emph{Searching for New Physics with Deep Autoencoders},
Phys. Rev. D 101 (2020) 075021. \href{https://arxiv.org/abs/1808.08992}{[arXiv:1808.08992]}

\bibitem{Cerri2019VAE}
O. Cerri, T.Q. Nguyen, F. Olness, D. Stoker et al.,
\emph{Variational Autoencoders for New Physics Mining at the Large Hadron Collider},
JHEP 05 (2019) 036. \href{https://arxiv.org/abs/1802.03426}{[arXiv:1802.03426]}

\bibitem{MetodievThaler2018JetTopics}
E.M. Metodiev, J. Thaler,
\emph{Jet Topics: Discrete Learning without Labels},
Phys. Rev. Lett. 120 (2018) 241602. \href{https://arxiv.org/abs/1802.00008}{[arXiv:1802.00008]}

\bibitem{Hallin2022CATHODE}
A. Hallin, J. Isaacson, G. Kasieczka, C. Krause, B. Nachman, T. Quadfasel, D. Shih, M. Sommerhalder,
\emph{Classifying Anomalies through Outer Density Estimation (CATHODE)},
Phys. Rev. D 106 (2022) 055006. \href{https://arxiv.org/abs/2109.00546}{[arXiv:2109.00546]}

\bibitem{Kasieczka2021LHCOlympics}
G. Kasieczka, B. Nachman, D. Shih et al.,
\emph{The LHC Olympics 2020: A Community Challenge for Anomaly Detection in High Energy Physics},
Rept. Prog. Phys. 84 (2021) 124201. \href{https://arxiv.org/abs/2101.08320}{[arXiv:2101.08320]}

\bibitem{Dillon2023INN}
B.M. Dillon, T. Plehn, C. Sauer, P. Wesslau,
\emph{Invertible Neural Networks for Collider Particle Physics},
Eur. Phys. J. C 83 (2023) 96. \href{https://arxiv.org/abs/2211.10612}{[arXiv:2211.10612]}

\bibitem{Brehmer2020Mining}
J. Brehmer, G. Louppe, J. Pavez, K. Cranmer,
\emph{Mining gold from implicit models to improve likelihood-free inference},
Proc. Nat. Acad. Sci. 117 (2020) 5242. \href{https://arxiv.org/abs/1805.12220}{[arXiv:1805.12220]}

\bibitem{Andreassen2019JUNIPR}
A. Andreassen, I. Feige, C. Frye, M.D. Schwartz,
\emph{JUNIPR: a Framework for Unsupervised Machine Learning in Particle Physics},
Eur. Phys. J. C 79 (2019) 102. \href{https://arxiv.org/abs/1804.09720}{[arXiv:1804.09720]}

\bibitem{Karagiorgi2022ML}
G. Karagiorgi, G. Kasieczka, S. Kravitz, B. Nachman, D. Shih,
\emph{Machine Learning in the Search for New Fundamental Physics},
SciPost Phys. 13 (2022) 082. \href{https://arxiv.org/abs/2112.03769}{[arXiv:2112.03769]}

\bibitem{Cacciari2008AntiKt}
M. Cacciari, G.P. Salam, G. Soyez,
\emph{The anti-$k_t$ jet clustering algorithm},
JHEP 04 (2008) 063. \href{https://arxiv.org/abs/0802.1189}{[arXiv:0802.1189]}

\bibitem{Cacciari2008Pileup}
M. Cacciari, G.P. Salam,
\emph{Pileup subtraction using jet areas},
Phys. Lett. B 659 (2008) 119. \href{https://arxiv.org/abs/0707.1378}{[arXiv:0707.1378]}

\bibitem{Salam2010Jetography}
G.P. Salam,
\emph{Towards Jetography},
Eur. Phys. J. C 67 (2010) 637. \href{https://arxiv.org/abs/0906.0777}{[arXiv:0906.0777]}

\bibitem{Thaler2011NSubjettiness}
J. Thaler, K. Van Tilburg,
\emph{Identifying Boosted Objects with N-subjettiness},
JHEP 03 (2011) 015. \href{https://arxiv.org/abs/1011.2268}{[arXiv:1011.2268]}

\bibitem{Larkoski2013ECF}
A.J. Larkoski, G.P. Salam, J. Thaler,
\emph{Energy Correlation Functions for Jet Substructure},
JHEP 06 (2013) 108. \href{https://arxiv.org/abs/1305.0007}{[arXiv:1305.0007]}

\bibitem{Larkoski2020JetSubstructureReview}
A.J. Larkoski, I. Moult, B. Nachman,
\emph{Jet Substructure at the Large Hadron Collider: A Review of Recent Advances in Theory and Machine Learning},
Phys. Rept. 841 (2020) 1. \href{https://arxiv.org/abs/1709.04464}{[arXiv:1709.04464]}

\bibitem{Marzani2019LookingInsideJets}
S. Marzani, G. Soyez, M. Spannowsky,
\emph{Looking inside jets: an introduction to jet substructure and boosted-object phenomenology},
Lect. Notes Phys. 958, Springer (2019). \href{https://arxiv.org/abs/1901.05390}{[arXiv:1901.05390]}

\bibitem{CMS2012DijetResonance}
CMS Collaboration (S. Chatrchyan et al.),
\emph{Search for resonances in the dijet mass spectrum from 7 TeV pp collisions at the LHC},
Phys. Lett. B 713 (2012) 40. \href{https://arxiv.org/abs/1206.1849}{[arXiv:1206.1849]}

\bibitem{CMS2012Compositeness}
CMS Collaboration (S. Chatrchyan et al.),
\emph{Search for quark compositeness in dijet angular distributions from pp collisions at 7 TeV},
Phys. Lett. B 718 (2012) 529. \href{https://arxiv.org/abs/1210.2387}{[arXiv:1210.2387]}

\bibitem{CMS2024DijetAnomaly}
CMS Collaboration,
\emph{Search for anomalous dijet events with the CMS detector at $\sqrt{s}=8$~TeV},
arXiv:2412.03747 (2024).

\bibitem{Brun1997ROOT}
R. Brun, F. Rademakers,
\emph{ROOT: An object oriented data analysis framework},
Nucl. Instrum. Meth. A 389 (1997) 81.

\bibitem{Pivarski2020Uproot}
J. Pivarski et al.,
\emph{Uproot: a Python package for reading and writing ROOT files},
Zenodo (2020). \href{https://doi.org/10.5281/zenodo.4340632}{doi:10.5281/zenodo.4340632}

\bibitem{Paszke2019PyTorch}
A. Paszke, S. Gross, F. Massa, A. Lerer et al.,
\emph{PyTorch: An Imperative Style, High-Performance Deep Learning Library},
Advances in Neural Information Processing Systems 32 (NeurIPS 2019). \href{https://arxiv.org/abs/1912.01703}{[arXiv:1912.01703]}

\bibitem{Pedregosa2011ScikitLearn}
F. Pedregosa, G. Varoquaux, A. Gramfort et al.,
\emph{Scikit-learn: Machine Learning in Python},
J. Mach. Learn. Res. 12 (2011) 2825.

\bibitem{Pujari2026Contrastive}
J.~J. Pujari, P.~A. Immadi, H.~Rasineni, T.~Bikku et al.,
\emph{Improving Stability of Discovery Significance in Higgs Boson Event Classification using Contrastive Representation Learning},
Discover Artificial Intelligence (2026). \href{https://doi.org/10.1007/s44163-026-01683-5}{doi:10.1007/s44163-026-01683-5}

\bibitem{Lipman2023FlowMatching}
Y. Lipman, R.T.Q. Chen, H. Ben-Hamu, M. Nickel, M. Le,
\emph{Flow Matching for Generative Modeling},
International Conference on Learning Representations (ICLR 2023). \href{https://arxiv.org/abs/2210.02747}{[arXiv:2210.02747]}

\bibitem{Chen2018NeuralODE}
R.T.Q. Chen, Y. Rubanova, J. Bettencourt, D. Duvenaud,
\emph{Neural Ordinary Differential Equations},
Advances in Neural Information Processing Systems 31 (NeurIPS 2018). \href{https://arxiv.org/abs/1806.07366}{[arXiv:1806.07366]}

\bibitem{Grathwohl2019FFJORD}
W. Grathwohl, R.T.Q. Chen, J. Bettencourt, I. Sutskever, D. Duvenaud,
\emph{FFJORD: Free-Form Continuous Dynamics for Scalable Reversible Generative Models},
International Conference on Learning Representations (ICLR 2019). \href{https://arxiv.org/abs/1810.01367}{[arXiv:1810.01367]}

\bibitem{Cowan2011Asymptotic}
G. Cowan, K. Cranmer, E. Gross, O. Vitells,
\emph{Asymptotic formulae for likelihood-based tests of new physics},
Eur. Phys. J. C 71 (2011) 1554. \href{https://arxiv.org/abs/1007.1727}{[arXiv:1007.1727]}

\bibitem{Rasineni2026MonoZNSF}
H.~Rasineni and B.~Chebrolu,
\emph{Mono-Z Dark Matter Search with Neural Spline Flows Using CMS Run 2015D Open Data},
arXiv:2607.13771 (2026). \href{https://arxiv.org/abs/2607.13771}{[arXiv:2607.13771]}

\bibitem{Rasineni2026HadronicFlowMatching}
H.~Rasineni and B.~Chebrolu,
\emph{Hadronic Mono-Z Dark Matter Sensitivity with Flow Matching on CMS Open Data},
arXiv:2609.02923 (2026). \href{https://arxiv.org/abs/2609.02923}{[arXiv:2609.02923]}

\end{thebibliography}
\end{document}